\documentclass[a4paper,fleqn,usenatbib]{mnras}

\usepackage[T1]{fontenc}
\usepackage{ae,aecompl}
\usepackage{xcolor}

\usepackage{graphicx}	
\usepackage{amssymb}	
\usepackage{pdflscape}	
\usepackage[english]{babel}

\newcommand{\msol}{M_{\rm \odot}}

\newcommand{\kms}{\rm{km \,s^{-1}}}
\newcommand{\kmsperkpc}{\rm{km\,s^{-1}\,kpc^{-1}}}
\newcommand{\rads}{\rm{rad \, s^{-1}}}

\title[Classifying spiral structures in simulations]{Classifying and modelling spiral structures in hydrodynamic simulations of astrophysical discs}
\author[D.H. Forgan, F.G. Ram\'on-Fox, I.A. Bonnell ]{D.H. Forgan $^{1,2}$\thanks{E-mail:dhf3@st-andrews.ac.uk}, F.G. Ram\'on-Fox$^{1}$ and I.A. Bonnell$^{1}$
\vspace{0.2cm} \\
$^{1}$SUPA, School of Physics and Astronomy, University of St Andrews, North Haugh, St Andrews KY16 9SS \\
$^{2}$St Andrews Centre for Exoplanet Science, University of St Andrews, St Andrews KY16 9SS \\
}

\begin{document}

\date{Submitted XXX Accepted YYY}

\pagerange{\pageref{firstpage}--\pageref{lastpage}} \pubyear{}

\maketitle

\label{firstpage}

\begin{abstract}

\noindent We demonstrate numerical techniques for automatic identification of individual spiral arms in hydrodynamic simulations of astrophysical discs.   Building on our earlier work, which used tensor classification to identify regions that were ``spiral-like'', we can now obtain fits to spirals for individual arm elements.  We show this process can even detect spirals in relatively flocculent spiral patterns, but  the resulting fits to logarithmic ``grand-design'' spirals are less robust.  Our methods not only permit the estimation of pitch angles, but also direct measurements of the spiral arm width and pattern speed.  In principle, our techniques will allow the tracking of material as it passes through an arm.  Our demonstration uses smoothed particle hydrodynamics simulations, but we stress that the method is suitable for any finite-element hydrodynamics system.  We anticipate our techniques will be essential to studies of star formation in disc galaxies, and attempts to find the origin of recently observed spiral structure in protostellar discs.

\end{abstract}


\section{Introduction}

\noindent Astrophysical disc structures are found across a wide range of scales -- from disc galaxies to discs surrounding active galactic nuclei (AGN), to discs around protostars and even around protoplanets.  Almost as commonly, these discs can be perturbed into producing spiral structures.

The means by which spiral structures are produced depends on the object being studied.  Broadly, spiral structure is caused by some form of unstable non-axisymmetric perturbation.  How this perturbation subsequently evolves into spiral structure depends on its origin, but in most cases is caused by the disc being susceptible to gravitational instability, where (for the gas) the Toomre parameter is

\begin{equation}
Q = \frac{c_s \kappa}{\pi G \Sigma} \lessapprox 1,
\end{equation}  

\noindent where $c_s$ is the sound speed of the gas, $\kappa$ is the epicyclic frequency, $G$ is the gravitational constant and $\Sigma$ is the local disc surface density.  A similar expression exists for the stability of a stellar disc (where the radial stellar velocity dispersion replaces $c_s$).  

Spiral structures have been observed in galaxies for some 150 years, going back to \citet{Rosse1850}'s observations of M51.  We now know that spiral galaxies constitute over half of observed massive galaxies \citep{Lintott2011}.  This has provided theorists with significant data reserves on which to test spiral generation theories.  

Spiral structures observed in disc galaxies can be self-generated, either by local instabilities/noise which are swing amplified to generate arms \citep[e.g. ][]{Sellwood1984}, or through globally propagating (quasi-stationary) density waves, as predicted by the pre-eminent Lin-Shu density wave theory \citep{Lin1964}.  Spiral structures can also be externally generated through tidal interactions, e.g. during galaxy mergers (\citealt{Holmberg1941}'s experiments in this area are particularly illuminating).  Indeed, all three mechanisms (instabilities, density waves, interactions) can and do collaborate to produce the spiral stuctures observed both in observations and simulations (see \citealt{Dobbs2014b} for a detailed review).

Spiral arms drive important physical processes in galaxies. Molecular clouds and recent star formation tend to be concentrated in and near spiral arms \citep{Schinnerer2013, Heyer2015,Ragan2016,Schinnerer2017}, most likely due to the compression and shocking of gas as it falls into the arm potential \citep{Bonnell2006,Dobbs2007,Bonnell2013}.  Spiral morphology is also directly correlated with the surface density of neutral atomic hydrogen in the disc, and the central stellar bulge mass \citep{Davis2015}.  

The gravitational torques induced by spiral structures will drive material into the inner few kpc of galaxies.   The potential asymmetries caused by stellar bars can then deliver this material further inward, feeding AGN (see e.g. \citealt{Alexander2012,Querejeta2016}). 

The wide-ranging effects of spirals on galaxy evolution has driven a great deal of effort on determining their properties in observations.  Attempts to characterise observed spiral structure in galaxies began with visual classification of the total number of arms, and their ``openness'' \citep{Hubble1926}, which can be quantified by the pitch angle $\phi$.  The modern Hubble sequence divides spiral galaxies into barred (SB) and unbarred (S) spirals, with a further sub-class (a-d) denoting the tightness of the spiral winding.  \citet{Elmegreen1990} proposed dividing spirals based on their arm number (grand design, multi-armed, and flocculent).  More quantitative measurements of observed spiral galaxies involve Fourier analysis of the image \citep{Rix1995,Foyle2010}, which decomposes the azimuthal variations in surface brightness into a Fourier sum over $m$ spiral modes with amplitude $A_m$, e.g.:

\begin{equation}
\frac{\mu(R,\theta)}{\bar{\mu}(R)} = \sum_{m=1}^{\infty} A_m (R) e^{im(\theta-\theta_m)}.
\end{equation}

\noindent Pitch angles can be fitted to images assuming a logarithmic spiral of constant $\phi$, a prediction of density wave theory \citep{Kennicutt1981}, or via hyperbolic spirals with radially varying pitch angle \citep[e.g.][]{Seiden1979}.  The pattern speed of an arm can be investigated by studying star formation (and tracers of star formation) inside the arm \citep{Egusa2004,Egusa2009}.  A steady arm of constant pattern speed will set up a sequence of tracers -- e.g. HI, CO, 24 $\mu$m emission from enshrouded stars, and UV emission from unobscured stars -- from upstream to downstream of the arm's corotation.  The failure to see this sequence in local spiral galaxies is strong evidence that spiral structures do not persist beyond the local dynamical time \citep{Foyle2011}.

Other observational techniques can also provide convincing evidence for the origin of spiral structures.  For example \citet{Choi2015} resolve the stellar populations along the northeast arm of M81, to show it does not possess a constant pattern speed.  This points to a kinematic origin, driven by tidal interactions with nearby galaxies M82 and NGC 3077s.


Protostellar discs can also self-generate spiral structure through local instabilities.  Shortly after the formation of a protostellar system, the protostellar disc has a mass comparable to that of the central star.  Such discs can soon assume a marginally unstable state where $Q\sim 1$, and the disc mass determines the spiral modes present \citep{Lodato2005,Forgan2011}.  Interactions with a companion either internal to the disc (such as a protoplanet), or external to it (such as a close stellar encounter) also generate tidally induced spiral arms. 

Spiral arms in protostellar discs are also efficient outward transporters of angular momentum, driving rapid accretion and assembly of protostars \citep{Laughlin1996}.  Spiral structures with a sufficient surface density contrast can concentrate dust grains \citep{Rice2004,Clarke2009,Booth2016}, promoting grain growth and setting the initial conditions for planet formation via core accretion, as well as altering local chemistry \citep{Ilee2011,Evans2015,Ilee2017}.

\noindent \emph{In extremis}, spiral arms of sufficiently large density amplitude can induce protostellar discs to fragment into bound objects \citep{Rice_et_al_05,Forgan2011a,Tobin2016}, providing an alternate formation channel for gas giants and substellar objects at large orbital semimajor axis \citep{TD_synthesis, TD_dynamics,Vigan2017}.  For sufficiently massive protostellar systems, fragmentation can also generate binary star systems \citep{Bonnell1994,Bate2002,Tobin2016}.

Spiral structure has only recently been observed in protoplanetary discs, initially in near-infrared (NIR) observations of scattered light, which is typically most sensitive to the upper disc surface.  In particular, extended two-armed structures have been detected around SAO206462 \citep{Muto2012} and MWC 758 \citep{Benisty2015}.  These arms are detectable out to relatively large distances (up to 100 au) from the parent star, with pitch angles of order $10^\circ$.  Recent ALMA observations have shown discs with spiral structure that extends down to the disc midplane, for example the recent detection of $m=2$ spiral structure around Elias 2-27, with a measured pitch angle of  7.9$\pm$0.4$^\circ$ \citep{Perez2016}  - the origin of this structure is not yet clear \citep{Meru2017}.  Characterising these spirals, and determining their origin, yields crucial information about protostellar accretion and the protostar's approach to the main sequence, as well as the formation of planetary systems.

If we are to identify the origin of newly observed protostellar disc spiral structures, or to study how spiral arms govern and drive star formation in galaxies, analysing spiral structure driven in hydrodynamic simulations is crucial.  As such, we require tools to identify and characterise spiral arms in these simulations.  Regardless of scale, characterising spiral morphology yields important diagnostics of what is driving the spiral structure.  In particular, the number of arms, their amplitude and pitch angles are sensitive to both the driving mechanism and the disc's properties.  For example, the spiral wake produced by planets embedded in a protostellar disc adopts a pitch angle which is a function of the disc temperature and rotation profile, as well as the planetary mass (\citealt{Rafikov2002}, see also \citealt{Zhu2015,Pohl2015}).  Gravitationally unstable discs also produce spiral structure with pitch angles and arm number that depend in particular on the disc mass \citep{Dong2015}, which may be a good deal larger than the observed disc mass \citep{Forgan2016e}.

What is clear is that both theoretical and observational astrophysicists stand to gain a great deal from higher quality characterisation of spiral structure in numerical simulations of astrophysical discs.  Theorists gain important insights into how spiral structures are generated, and how they affect the thermal and chemical history of gas and dust, and the future evolution of the disc.  Observers gain diagnostics for what is driving the spiral structure in their observations, and glean information on disc properties that is generally orthogonal to other methods.

Most attempts to characterise spiral structure in simulations rely on Fourier decomposition (e.g. \citealt{Cossins2008, Dobbs2010,Forgan2011,Mata-Chavez2014,Pettitt2016}).  This gives important information on the relative strengths of the spiral modes at play, and the pitch angle and pattern speed of the dominant mode.  However, it does not give the pitch angle and pattern speed of individual arms.  It also does not inform us as to what sections of the disc are currently in the spiral (and which sections are in the interarm regions).

Recent attempts at spiral arm characterisation have moved away from this Fourier analysis.  For example, \citet{Grand2012} attempted to directly identify individual stellar spiral arms in N-Body/hydrodynamic simulations of a barred spiral galaxy, by finding the location of a series of density peaks in a range of annuli. However, this does not directly provide data on the arm width, or which fluid elements currently reside in the spiral.

In this paper, we demonstrate that judicious use of tensor classification on hydrodynamic simulation data \citep{Forgan2016} allows the identification of fluid elements that are inside spiral structures (or in the interarm regions).  Further analysis allows the isolation of individual arms to obtain their shape parameters, as well as the pattern speed of the wave. Our examples focus on smoothed particle hydrodynamics (SPH) simulations, but we emphasise that our methods only require the ability to compute derivatives, and are therefore applicable to any hydrodynamic simulation.




\section{Methods }\label{sec:methods}

\noindent Our spiral detection algorithm has two distinct parts.  In the first part, fluid elements undergo tensor classification \citep{Forgan2016} to determine whether their behaviour indicates they are in fact inside a spiral structure.  In the second part, the fluid elements identified as spirals are extracted from the main simulation, and a friends-of-friends algorithm is used on this population to identify the spine of each individual spiral.  We describe these procedures below.  Our code is published on Github at \url{https://github.com/dh4gan/tache}.

\subsection{Tensor Classification}

\noindent We follow the same procedure as described in \citet{Forgan2016}, which itself builds on work originally applied to $N$-Body simulations of the cosmic Web \citep[see e.g.][]{Hahn2007,Forero-Romero2009}.  In our formalism, tensor classification determines the topology of a chosen field at the location of a given SPH particle.  We consider the topology of either the gravitational potential $\Phi$, by computing the tidal tensor $T_{ij}$:

\begin{equation}
T_{ij} = \frac{\partial^2 \Phi }{ \partial x_i \partial x_j},
\end{equation}

\noindent or the velocity field via the velocity shear tensor $\sigma_{ij}$:

\begin{equation}
\sigma_{ij} = -\frac{1}{2}\left(\frac{\partial v_i}{\partial x_j} + \frac{\partial v_j}{\partial x_i} \right).
\end{equation}

\noindent The classification proceeds as follows.  Once the tensor to be used ($T$ or $\sigma$) is selected, its eigenvalues $\lambda_i$ and their corresponding eigenvectors $\mathbf{n_i}$ are computed for every particle, e.g.:

\begin{equation}
T \mathbf{n}_j = \lambda_j \mathbf{n}_j
\end{equation}

\noindent The eigenvalues are defined so $\lambda_1 \geq \lambda_2 \geq \lambda_3$.  As $T$ and $\sigma$ are real and symmetric, these eigenvalues are always real.  Both tensors assume that the fields being investigated (potential, velocity) are smooth and continuous, so that the derivatives are always defined.  

\noindent We then compute $E$, which is defined as the number of positive eigenvalues\footnote{In practice, $E$ is defined as the number of particles whose eigenvalues exceed a small, non-zero threshold, see \citet{Forgan2016} for details.}.  The value of $E$ determines a topological classification for each fluid element:

\begin{itemize}
\item $E=0 \rightarrow $ ``void'' (0-D manifold)
\item $E=1 \rightarrow $ ``sheet'' (1-D manifold)
\item $E=2 \rightarrow $ ``filament'' (2-D manifold)
\item $E=3 \rightarrow $ ``cluster'' (3-D manifold)
\end{itemize}

\noindent We can see this by considering the tidal tensor in the context of Zeldovich theory \citep{Zeldovich1970}.  A test particle orbiting a local extremum $\nabla \Phi = 0$ has the following (linearised) equation of motion:

\begin{equation}
\ddot{x}_{i} = -T_{ij} x_j.
\end{equation}

\noindent If we operate in a basis where $T$ is diagonal, then the stability of the orbit is determined by the sign of the eigenvalues $\lambda_i$.  A single positive eigenvalue ensures a stable orbit along the corresponding axis (or a sheetlike structure, if multiple particles are present).  Two positive eigenvalues allow an extra degree of freedom (filaments) and finally all three being positive allow orbits of any degree (clusters).  If no eigenvalues are positive, stable orbits cannot be achieved (voids).

Tensor classification using the tidal tensor therefore determines the manifold dimension of the local potential, and as such the preferred structure for matter to collapse into if only gravity is present.  Classification using the velocity shear tensor instead diagnoses the manifold being sculpted by the flow at any given instant.

\noindent As a result, the finite elements of any hydrodynamic simulation can be grouped into four components, with each component composed of fluid elements of a matching $E$.  Discs are inherently sheet-like structures, and hence perturbations from axisymmetry are classified as either filaments or clusters.  We will see that depending on the strength of spiral structures, fluid elements classified as either filaments or clusters trace spiral structures, and that fluid elements classified as sheets will trace unperturbed disc material in the interarm regions.  As a general rule, the tidal tensor is better suited to tracing high-amplitude spiral structures that induce a strong surface density perturbation, but the velocity shear tensor is better at revealing low-amplitude spiral structures that produce weak surface density perturbations\footnote{Tensor classification is designed to operate on smooth continuous fields.  This is not a concern as SPH fields are guaranteed to be smooth, but it is unclear how strong discontinuities will affect tensor analysis.  If reliable derivatives of a field cannot be computed for a simulation region, that region cannot be reliably classified.}

Tensor classification also yields eigenvectors corresponding to each eigenvalue.  In sheets, the eigenvector $\mathbf{n}_1$ corresponding to the single \emph{positive} eigenvalue $\lambda_1$ defines the normal or symmetry axis of the sheet.  In filaments, the eigenvector $\mathbf{n}_3$ corresponding to the single \emph{negative} eigenvalue $\lambda_3$ provides the flow direction of the filament.  We do not use eigenvectors in this analysis, but note that they may also be of interest in e.g. determining the relative geometry of spiral structures to their host disc.

\subsection{Identifying individual spirals}

Once tensor classification is complete, we extract all particles that correspond to either filament or cluster classification, as these trace the spiral structure in the disc, and discard the other particles.  To determine the spines of each spiral arm on this subset, we run a friends-of-friends algorithm as follows.

We specify a minimum spherical radius from the centre of the disc, $r_{\rm min}$, from which to begin our study.  As we are interested in spines, we will also select the top $x^\%$ percentile in density of the remaining particles.

We then select particle $i$ with the largest density, where radius $r_i > r_{\rm min}$.  This particle forms the first component of the spiral's spine.

A sphere of radius $L$ is then drawn from $r_i$.  The particles inside the sphere are tested, and particle $j$ is selected if it is the densest particle with $r_j > r_i$.  Particle $j$ then forms the second component of the spiral's spine, $j$ is then set to $i$, and the algorithm is repeated until a particle can no longer be found that meets the above conditions.  Whenever a particle is tested, it is removed from the list of particles and is not tested again, regardless of whether it forms a component of the spiral spine or otherwise.  The act of testing assigns the particle to that particular spiral.  As a consequence, any particle within a distance $L$ of any spiral spine point belongs to that spiral.

If the conditions can no longer be met, a new spiral is begun at the location of the densest particle not yet tested, and the procedure begins again.  Note that throughout we are implicitly assuming that the particle density decreases with increasing radius.

Three parameters form the identification algorithm: the linking length $L$, the minimum radius from which to begin, $r_{\rm min}$, and the percentile of the population from which to conduct the analysis, $x^\%$.

The values needed for these three parameters will depend on the resolution and input physics of the simulation being analysed.  The linking length $L$ should clearly be smaller than the spiral arm spacing to avoid the algorithm ``jumping'' between individual arms, while still being larger than the typical smoothing length to obtain a sufficient number of neighbouring particles.  The minimum radius should be selected based on prior knowledge of where spiral structure is well-resolved (as well as the locations from which the user desires to measure said structure).  

The percentile selection depends on how well-defined the structures are post-tensor classification.  Selecting a low $x^\%$ will result in easier classification, but will prevent the classification of weak structures. 

This algorithm delivers $N_{\rm spiral}$ sets of co-ordinate points, with each set denoting the spine of a spiral structure.  We elect to fit these points via Markov Chain Monte Carlo (MCMC) to a logarithmic spiral:

\begin{equation}
r = r_0 + a e^{b\theta},
\end{equation}

\noindent with fit parameters $(r_0= \sqrt{x_0^2 + y^2_0}, a,b)$.  The origin of the spiral is defined at $\theta=0$; $r=r_0 +a$.  A pure logarithmic spiral will have a constant pitch angle $\phi$, which in radians is simply

\begin{equation}
\phi = \arctan b.
\end{equation}

It is worth noting that if the spiral arms exclusively constitute the densest regions of our simulation, then it is likely that tensor classification will not be necessary to determine the location of the spiral spines, and that the above friends-of-friends algorithm will work perfectly well.  However, if there are other dense structures in the simulation (such as condensing clumps), then the algorithm will attempt to erroneously fit spiral spines to them.  We have run test calculations on fragmenting discs to show that tensor classification can successfully remove dense structures that are not spiral arms, preventing this problem.  Also, without tensor classification, only the spine of the arm can be determined -- the width of the spiral arms, or the properties of the gas contained within the arm cannot be measured.

\section{Tests}

\noindent We now test our algorithm on three examples of spiral structure in astrophysical discs.  The origins of the structure in each case are subtly different.  In the first, we consider a self-gravitating protostellar disc, whose spiral structure is entirely due to the self-gravity of the disc gas.  In the second case, we consider a spiral galaxy with an imposed fixed potential that induces well-defined structures.  In the last, we consider a galaxy with a ``live potential'' of star particles, with a resulting spiral structure that is highly flocculent.

We use smoothed particle hydrodynamics (SPH) for all three tests.  SPH is a Lagrangian hydrodynamics method, where the fluid is discretised into individual particles.  Each SPH particle $i$ possesses position and velocity vectors $\mathbf{r}_i$, $\mathbf{v}_i$, internal energy $u_i$ and smoothing length $h_i$, which is adjusted such that a sphere of radius $2h_i$ encompasses a suitable number $N$ of neighbour SPH particles.

The local fluid density is established by performing a sum over the neighbour particles' smoothing kernels:

\begin{equation}
\rho (\mathbf{r}) = \sum^{N}_{i=1} m_i W(\left|\mathbf{r} - \mathbf{r}_i\right|, h). \label{eq:sph-rho}
\end{equation}

\noindent Writing down the system's Lagrangian and applying variational methods (in combination with equation \ref{eq:sph-rho}) produces an algorithm for solving the equations of hydrodynamics on the particle distribution (with appropriate compensation for shocks and fluid mixing, see \citealt{Monaghan_92,Monaghan_05,Price2012} for reviews).

Table \ref{table:fit-params} shows the values of $L$, $r_{\rm min}$ and $x^\%$ used in the three tests run in this paper.

\begin{table}
	\centering
	\caption{The fit parameters used in the spiral detection algorithm for the test cases evaluated in this paper.}
	\begin{center}
		\begin{tabular}{ccccc}
			\hline
			Section & Simulation & $L$ & $r_{\rm min}$ & $x^\%$  \\
			\hline
			3.1  & Protostellar disc & 6 au & 10 au & 20 \\
			3.2 & Galaxy, analytic potential & 0.2 kpc & 1 kpc & 30 \\
			3.3 & Galaxy, live potential & 0.2 kpc & 1 kpc & 30 \\ 
			\hline
		\end{tabular}
	\label{table:fit-params}
	\end{center}
\end{table}

\subsection{A self-gravitating protostellar disc}

\begin{figure*}
\begin{center}$\begin{array}{cc}
\includegraphics[scale=0.3]{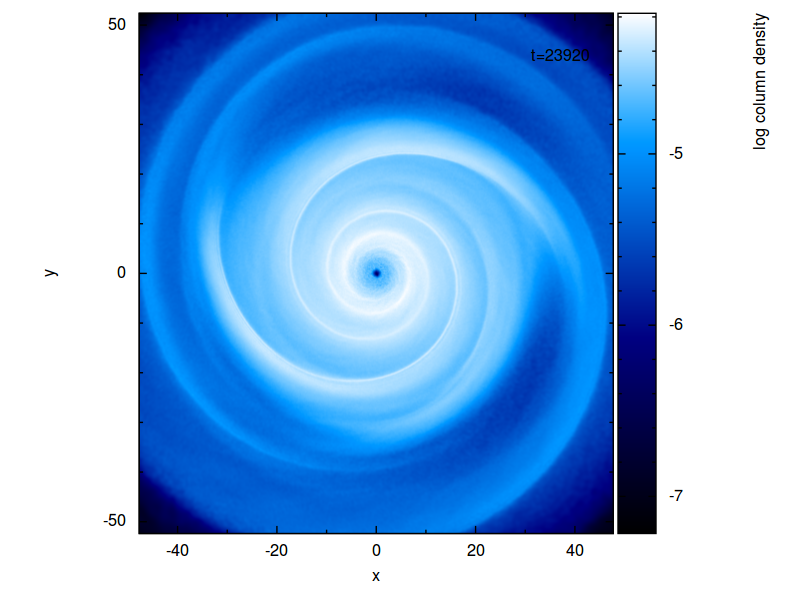} &
\includegraphics[scale=0.3]{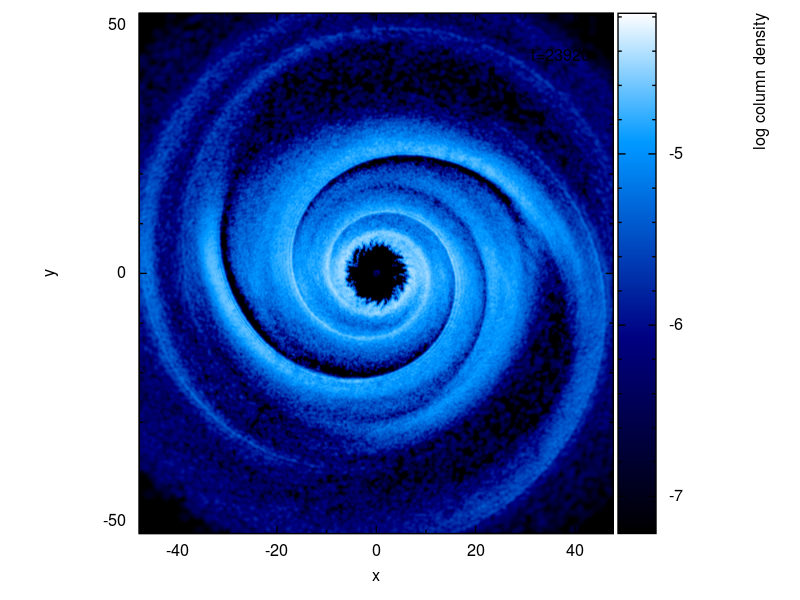} \\
\includegraphics[scale=0.4]{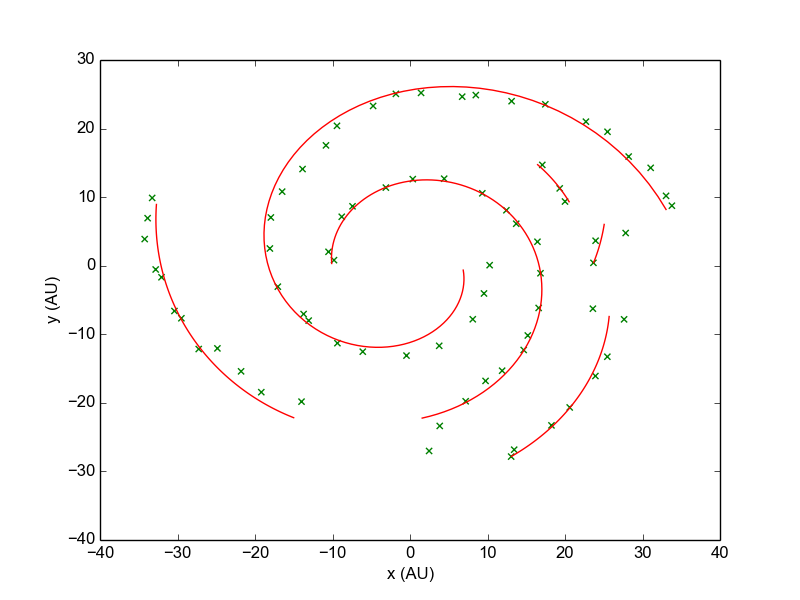} &
\includegraphics[scale=0.3]{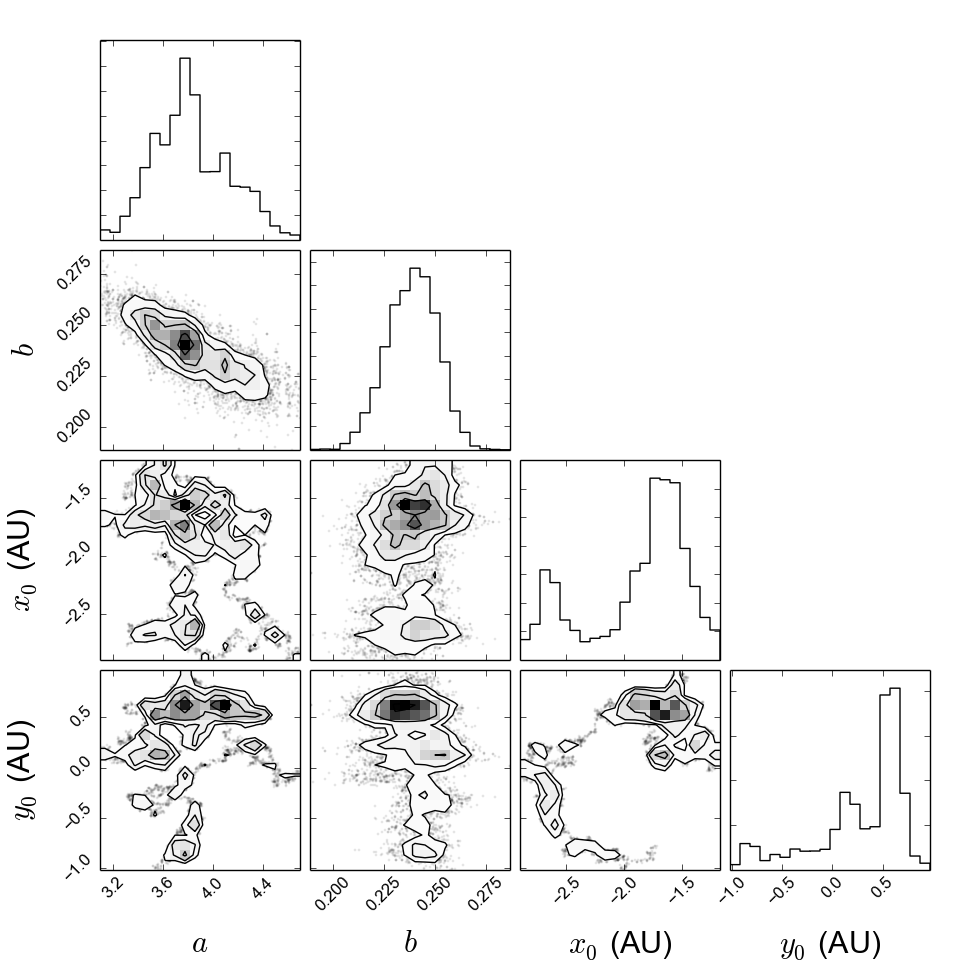}
\end{array}$
\end{center}
\caption{Spiral arm detection in a self-gravitating protostellar disc.  Top left: a snapshot from the full SPH simulation.  Top right: the same snapshot, with only the SPH particles identified as exhibiting arm-like behaviour (through classification of the tidal tensor).  Bottom left: the spine points of each arm (green crosses) accompanied by the best fit logarithmic spirals in red.  Bottom right: The posterior distribution of the logarithmic spiral parameters for the uppermost spiral. \label{fig:sgdisc}}
\end{figure*}

\begin{figure*}
\begin{center}$\begin{array}{cc}
\includegraphics[scale=0.4]{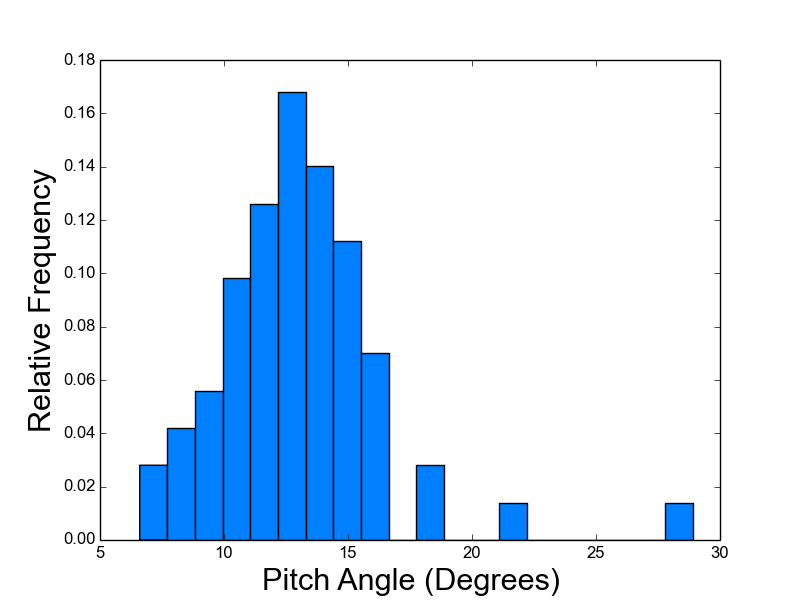} &
\includegraphics[scale=0.4]{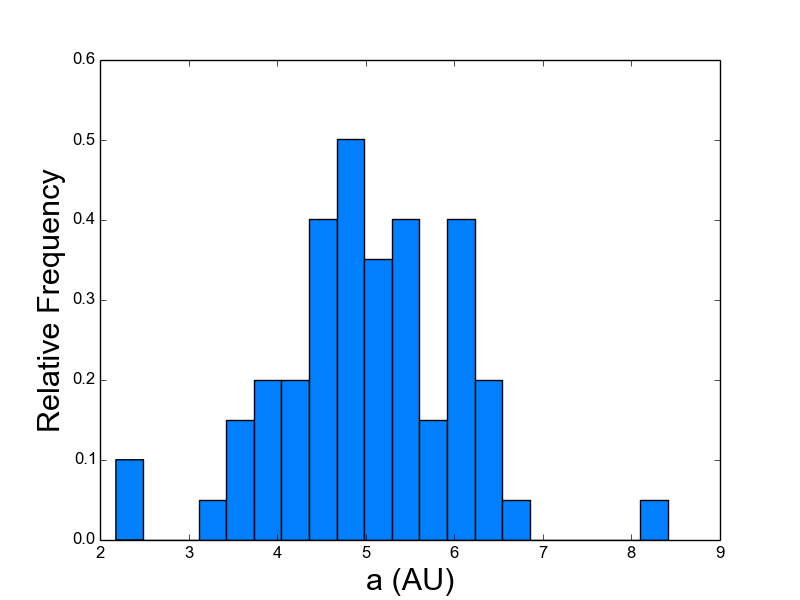} \\
\end{array}$
\end{center}
\caption{The distribution of pitch angle $\phi$ (left) and $a$ over the course of 40 snapshots of the self-gravitating disc simulation (approximately 500 years of evolution). \label{fig:sgdisc_histograms}}
\end{figure*}

\begin{figure}
\begin{center}
\includegraphics[scale=0.4]{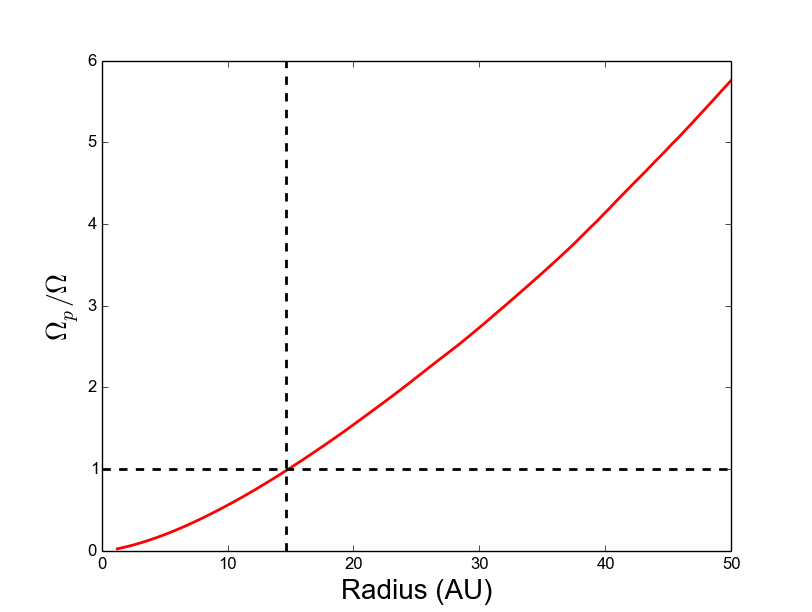}
\end{center}
\caption{The ratio of the pattern speed to the bulk angular velocity $\Omega_p/\Omega$ as a function of radius.  Corotation occurs when this ratio is unity, which occurs at 14.6 au, interior to which the disc is dominated by artificial viscosity. \label{fig:sgdisc_corotation}}
\end{figure}

\noindent For our first test, we return to a simulation previously analysed in \citet{Forgan2016}.  The top left panel of Figure \ref{fig:sgdisc} shows an SPH simulation of a self-gravitating disc of mass $0.25 \msol$, orbiting a sink particle representing a star of $1\msol$.  The initial disc surface density profile follows $\Sigma \propto r^{-1}$, and has a maximum radius of 50 au.  The sound speed profile fixes the initial Toomre parameter $Q=2$ at all radii.

The radiative transfer formalism of \citet{intro_hybrid} is used, implementing realistic local cooling alongside typical compressive and shock heating to settle into a self-regulated, marginally stable state, where the Toomre parameter $Q\sim 1.5$.

Marginally stable self-gravitating discs are able to sustain quasi-steady spiral structure for as long as the Toomre parameter can be held at this value, which depends critically on the disc's total mass, and its ability to cool.  The top right panel of Figure \ref{fig:sgdisc} shows the result of tensor classification using the tidal tensor.  Given the deep gravitational perturbations caused by this spiral structure, we elect to use $E=3$ (cluster) particles to define the minimum potential of the structure.

The bottom row of Figure \ref{fig:sgdisc} shows the result of spiral arm identification.  Bottom left shows the identified spine points of each spiral (green points), with accompanying logarithmic fits via MCMC (red curves).  An example of the derived posteriors from the MCMC fits is shown in the bottom right panel (for the uppermost spiral).  

We see that logarithmic spirals can produce good quality fits to the data, especially in the outer disc.  The uppermost spiral is well fitted by a logarithmic spiral with $(a,b) = (3.634 \pm 0.300,0.2468 \pm 0.0123)$, giving a pitch angle of $13.8^\circ \pm 0.6 ^\circ$.   The posterior distributions for $a$ and $b$ are well behaved and unimodal, with only a weak degeneracy between $a$ and $b$.

The inner radii of the spiral shows some deviation from the fit, suggesting that pitch angles may not be constant in this region.  This is reflected in the best fit centre of the spiral deviating by around 1-2 au from the star's actual position at the origin.

Repeating this analysis over a subsequent 40 timesteps of the simulation (around 1.5 outer rotation periods, Figure \ref{fig:sgdisc_histograms}) shows that the matched spirals tend to share relatively similar parameters over this interval.  The sample mean pitch angle derived over this duration is $13.02^\circ$, with a sample standard deviation of around $3.4^\circ$.  Interestingly, the scale parameter $a$ tends to assume slightly larger values than our example spiral, with a mean of 4.996 au, and standard deviation 1.02 au.

Being able to measure the spiral structure over more than one timestep allows us to calculate the pattern speed of a given arm.  Given that for a logarithmic spiral

\begin{equation}
\theta = \frac{1}{b} \ln (r/a),
\end{equation}

\noindent we can estimate the pattern speed $\Omega_p = \dot{\theta}$ given $\dot{b}$ and $\dot{a}$:

\begin{equation}
\dot{\theta} = \frac{-\dot{b}}{b} \theta -\frac{\dot{a}}{ab}.
\end{equation}

\noindent Note that $\dot{\theta}$ has a dependence on $\theta$ only if the pitch angle is time-dependent.  Having computed $\dot{a}$ and $\dot{b}$ from the two fits, we can evaluate the pattern speed over the range $\theta=[0,2\pi]$ and take an average.

\noindent For our example spiral, we evaluate $(a,b)$ for two snapshots and compute $\Omega_p = 3.5 \times 10^{-9}\, \rads$.  The corotation radius, where $\Omega=\Omega_p$, is approximately 14.6 au (Figure \ref{fig:sgdisc_corotation}).  

Artificial viscosity dominates the disc interior to this radius (see e.g. \citealt{Artymowicz1994,Murray1996,Lodato_and_Rice_04, Lodato2010, Forgan2011}). The Toomre $Q$ parameter also reaches the instability regime just beyond this radius ($\sim 20$ au).  As a result, the total disc stress has an artificial minimum at corotation, rather than continuing to decrease with decreasing radius \citep{Rice_and_Armitage_09}.

This shows how the inner disc resolution affects the launching point of spiral structures.  Indeed, unstable non-axisymmetric normal modes in self-gravitating gaseous discs are expected to trail the flow at all radii \citep{Papaloizou1991}.  


\subsection{A disc galaxy with fixed arm potential}

\begin{figure*}
\begin{center}$\begin{array}{cc}
\includegraphics[scale=0.3]{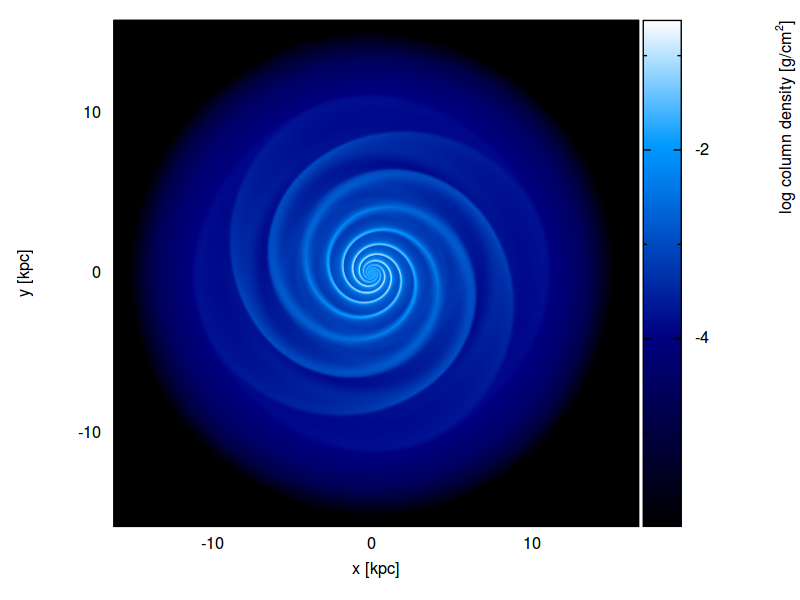} &
\includegraphics[scale=0.3]{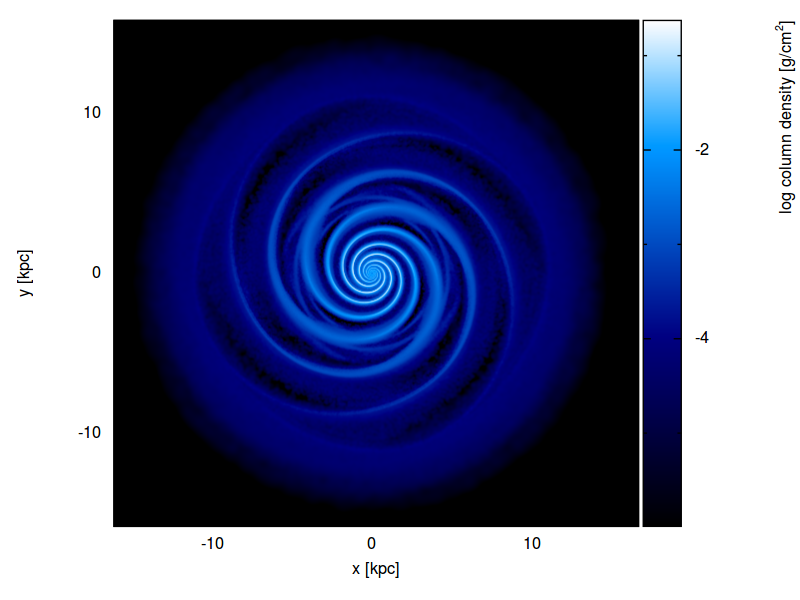} \\
\includegraphics[scale=0.3]{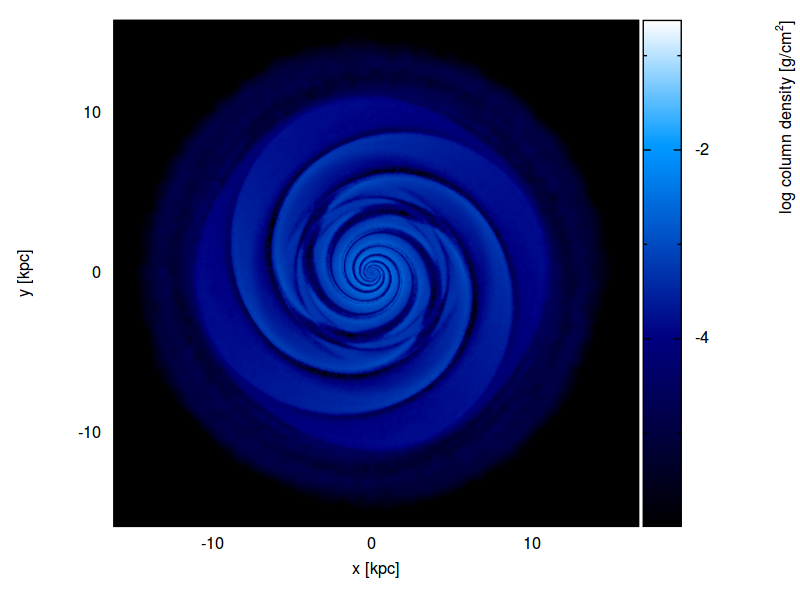} &
\includegraphics[scale=0.4]{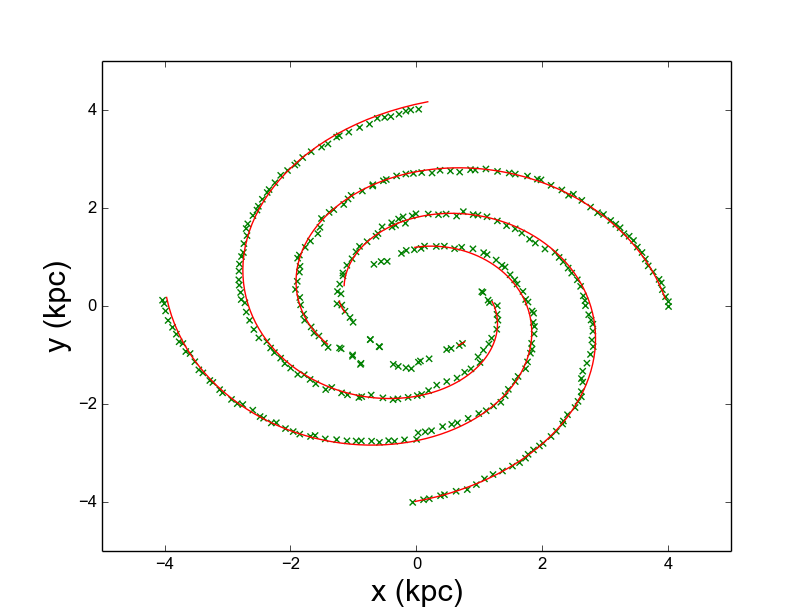} \\
\end{array}$
\end{center}
\caption{The evolution of a galactic disc under an analytic spiral potential.  Top left: the full SPH simulation containing 2m gas particles. Top right: the particles identified as ``spiral-like'' (filament class).  Bottom left: particles identified as moving in a disc configuration (sheet class).  Bottom right: Fits to the spiral structure.\label{fig:fixedgalaxy}}
\end{figure*}

\begin{figure}
\begin{center}
\includegraphics[scale=0.4]{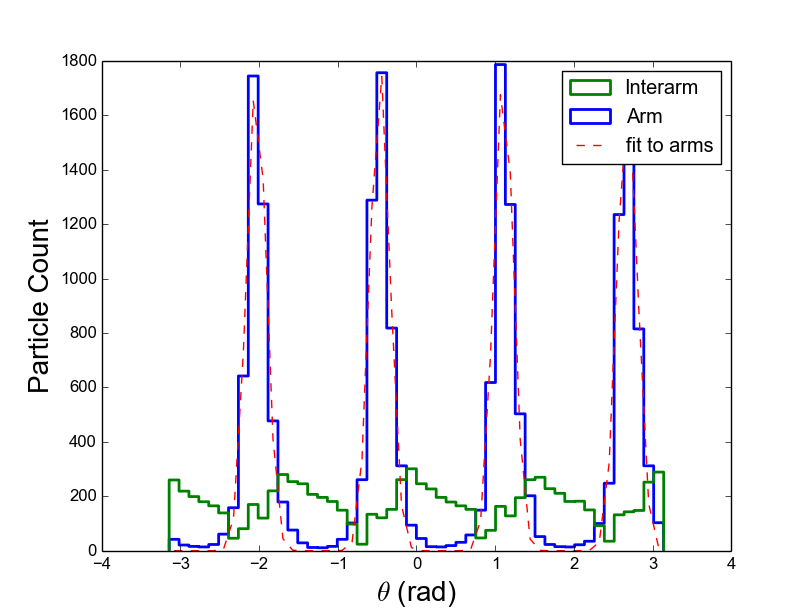}
\end{center}
\caption{The number of particles classified as ``arm'' vs ``interarm'' in the galactic disc under the analytic potential, as a function of $\theta$, in an annulus with inner and outer radii $R=1.95, 2.05$ kpc.  Also plotted are Gaussian fits to the arm histograms, with each Gaussian possessing a full width half maximum (FWHM) of 0.2825 rad.  The galaxy rotates in the positive $\theta$ direction. \label{fig:fixedgalaxy_theta}}
\end{figure}

\begin{figure*}
\begin{center}$\begin{array}{cc}
\includegraphics[scale=0.4]{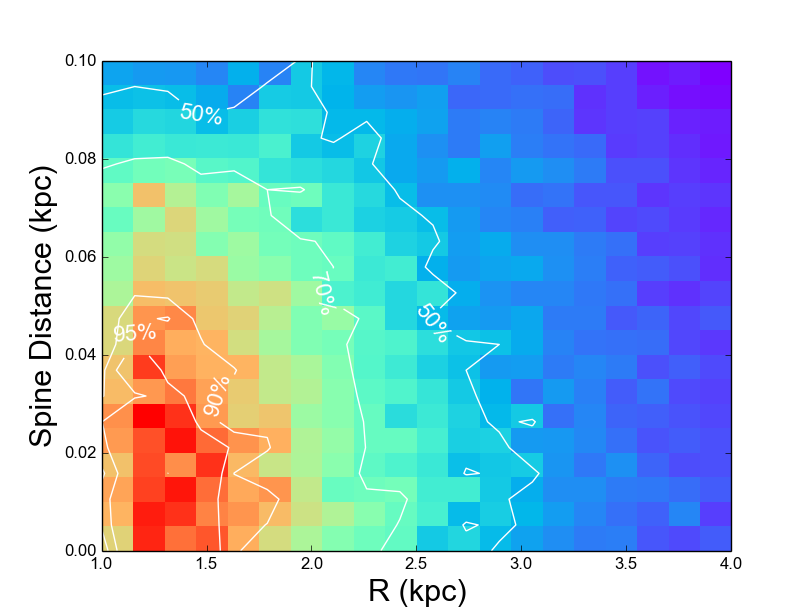} &
\includegraphics[scale=0.4]{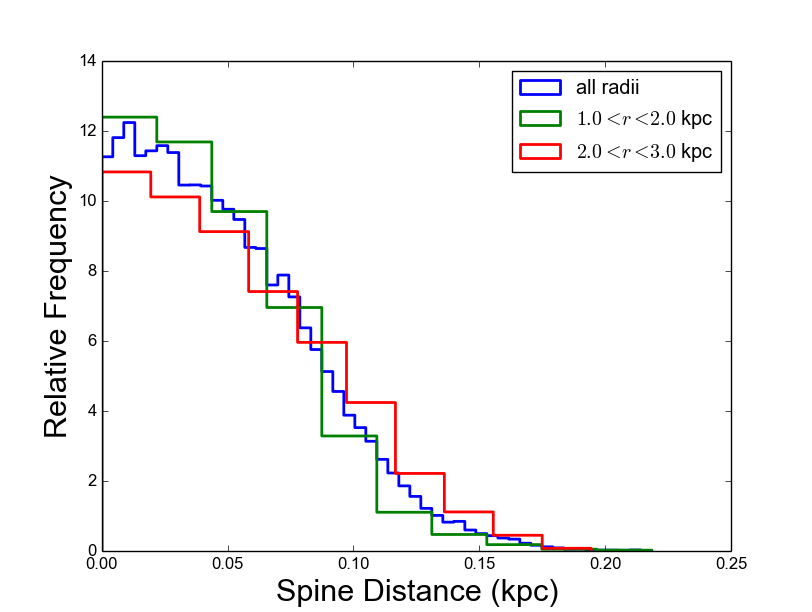} \\
\end{array}$
\end{center}
\caption{The width of an individual spiral arm in the galaxy model with analytic potential, as a function of radius.  Left: For a population of particles identified by our algorithms as belonging to a specific arm, we compute their minimum distance from the arm and bin them in the above 2D histogram.  We overplot percentiles of the bin counts to highlight how the typical separation from an arm is constant with radius. Right: 1D histograms of distance from the arm for the same population, for all radii, and inner and outer radii.  \label{fig:fixedgalaxy_width}}
\end{figure*}

In our second test, we use a galaxy model with analytically defined spiral structure imposed by an external gravitational potential. The galactic potential is represented by a combination of an axisymmetric term plus a perturbation of the spiral arms: 

\begin{equation}
\Phi = \Phi_{0} + \Phi_{pert}
\end{equation}

The first component is given by a logarithmic potential (e.~g. \citealt{BandT2008}):

\begin{equation}
\Phi_0 = \frac{1}{2} v^2_0 \left(R^2 + R^2_c + (z/z_q)^2\right),
\end{equation}

\noindent which has a rotation curve given by:

\begin{equation}
	v_c(R) = v_0 \frac{R}{\sqrt{R_c^2 + R^2}},
	\label{eq:rot-curve}
\end{equation}
where $v_0$ is a velocity parameter, $R_c$ is a characteristic radius, and $z_q$ is a vertical scale factor. This produces a flat rotation curve at values larger than $R_c$. For our simulations, $v_0 = 120 \,\kms$, $R_c = 2.5$ kpc, and $z_q = 0.7$. 

The velocity parameter corresponds to that of the rotation curve of the local spiral galaxy M33 \citep{CorbelliSalucci2000,Seigar2011,Kametal2015}. The spiral arm perturbation uses the scheme of \citet{CoxandGomez2002}:

\begin{equation}
\Phi_{pert}(R,\theta,z, t) = \sum_n A_n(R,z) \cos \left(n \Gamma(R,\theta,t) \right),
\end{equation}

\noindent where $A_n$ describes the perturbation amplitude which decays with both $R$ and $z$ (see \citealt{CoxandGomez2002} for details).  $\Gamma$ determines the form of spiral generated by the perturbation:

\begin{equation}
\Gamma(R,\theta,t) = N \left(\theta + \Omega_p t - \frac{\ln(R/R_0)}{\tan \phi} - \theta_p \right).
\end{equation}

\noindent The number of spiral arms $N = 4$, and their pattern speed is $\Omega_p = 23 \,\kmsperkpc$ (the pitch angle $\phi$ is 15$^{\circ}$). The constant $\theta_p$ defines the location of the arm at a fiducial radius $R_0$.

The gaseous component is assumed to have an exponential surface density profile: $\Sigma(R) = \Sigma_0 e^{-R/R_d}$, where $\Sigma_0$ is the central surface density and $R_d$ is the scale radius. The gas self-gravity is not included. 

This analytic potential model is advantageous for its reduced computational expense, and also allows us to have a well defined spiral structure for testing the spiral detection algorithm.

Figure \ref{fig:fixedgalaxy} shows the full SPH simulation (top left), the particles identified as ``spiral-like'' using the velocity shear tensor (top right), and the particles identified as belonging to the interarm regions ($E=1$, ``sheet'' classification, bottom left).  The spines of the four main spirals are easily identified (bottom right of Figure \ref{fig:fixedgalaxy}), and are well fitted by logarithmic spirals each with $b\approx 0.25 \pm 5\%$ (i.e. $\phi \approx 14^\circ \pm 0.7^\circ$),  representing the input perturbation model well.  

The separation of arm and interarm particles in azimuth is clear, if we bin particles by class in an annulus of 0.1 kpc (Figure \ref{fig:fixedgalaxy_theta}). We can in fact apply Gaussian fits to the four arms to determine their angular width, resulting in a full width half maximum (FWHM) of 0.2825 radians.  This corresponds to a rather large distance of around 0.5 kpc at this radius, but this is principally due to the winding of the arms aligning them with the azimuthal vector.

We can compute the width of a spiral arm along its entire extent by interrogating the particles that have been identified as belonging to it.  We do this for the rightmost spiral in the bottom right panel of Figure \ref{fig:fixedgalaxy}, and compute each particle's minimum distance to the spiral spine.  The left panel of Figure \ref{fig:fixedgalaxy_width} shows a 2D histogram of the data (we overplot percentiles of the bin counts for convenience).  We also extract 1D histograms by binning in radius (right panel of Figure \ref{fig:fixedgalaxy_width}) to show that the spiral arm width is essentially constant with radius (with the FWHM of the 1D histograms being approximately 0.1 kpc).

As for the protostellar disc, we can compute pattern speeds by comparing two snapshots of the simulation.  We find pattern speeds of around $20.4\, \kmsperkpc$ ($6.6 \times 10^{-16} \mathrm{rad \, s^{-1}}$), with a $1-\sigma$ uncertainty of around 19\%.  Given the  rotation curve, these parameters define the corotation radius as

\begin{equation}
R_{\rm co} = \sqrt{\left(\frac{v_0}{\Omega_p}\right)^2 - R^2_c} = 5.33 \, \rm{kpc}
\end{equation}

\noindent The analytic spiral arm potential has a pattern speed of $23\, \kmsperkpc$ ($7.45\times 10^{-16} \mathrm{rad \, s^{-1}}$), and corotation at 4.9 kpc.  Hence, the analytic value resides within the $1-\sigma$ uncertainties of our calculation.


\subsection{A disc galaxy with a live stellar potential}

\begin{figure*}
\begin{center}$\begin{array}{cc}
\includegraphics[scale=0.3]{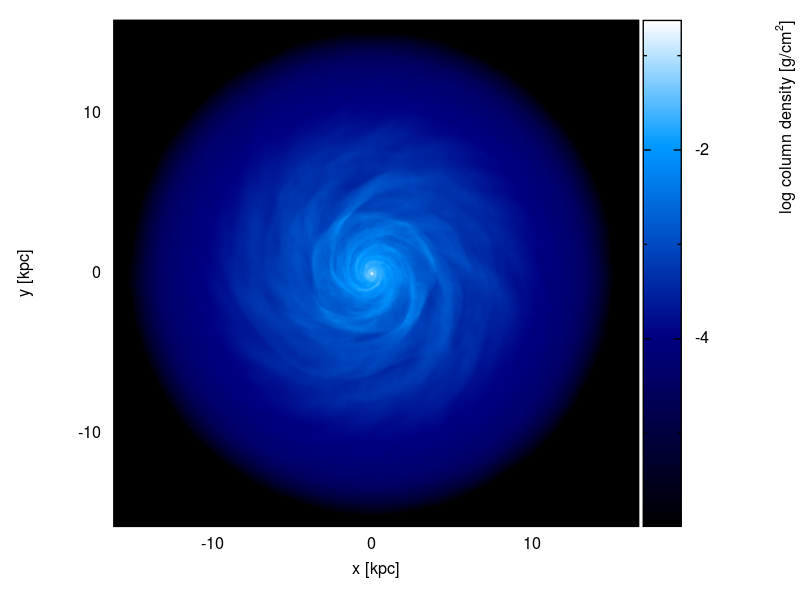} &
\includegraphics[scale=0.3]{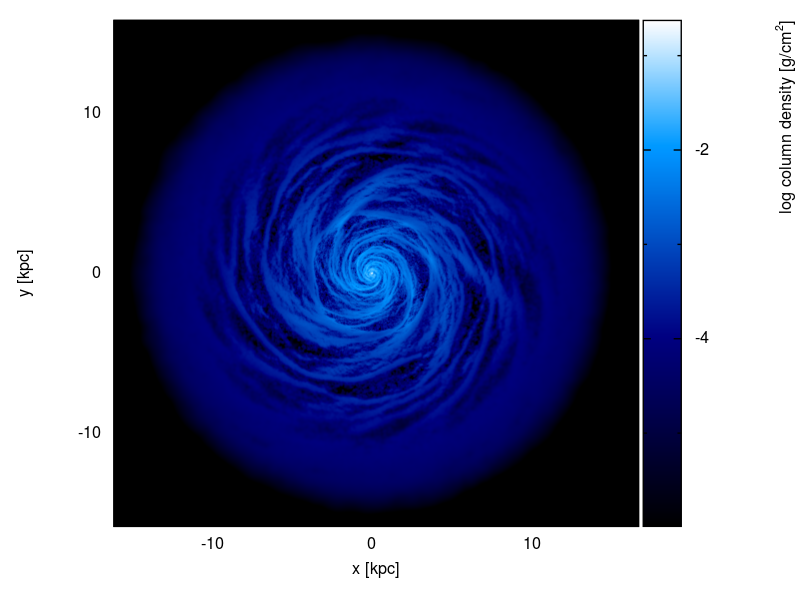} \\
\includegraphics[scale=0.3]{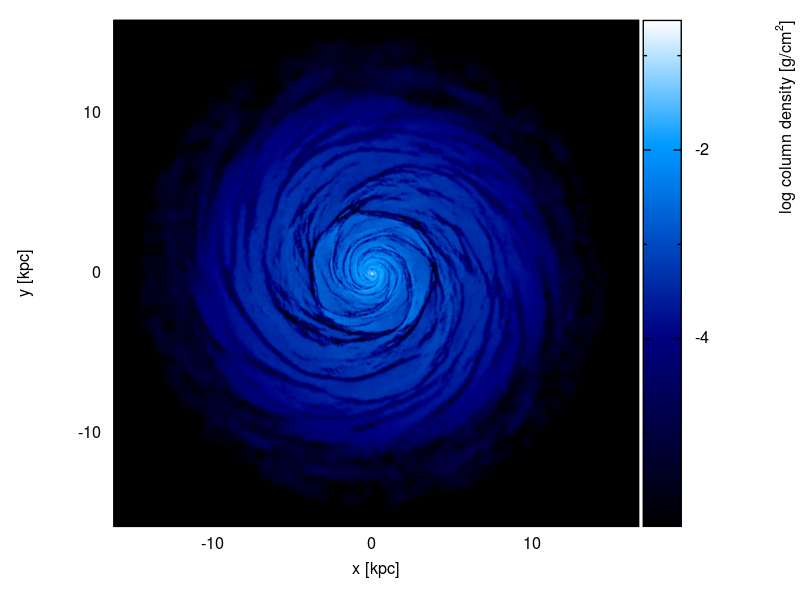} &
\includegraphics[scale=0.25]{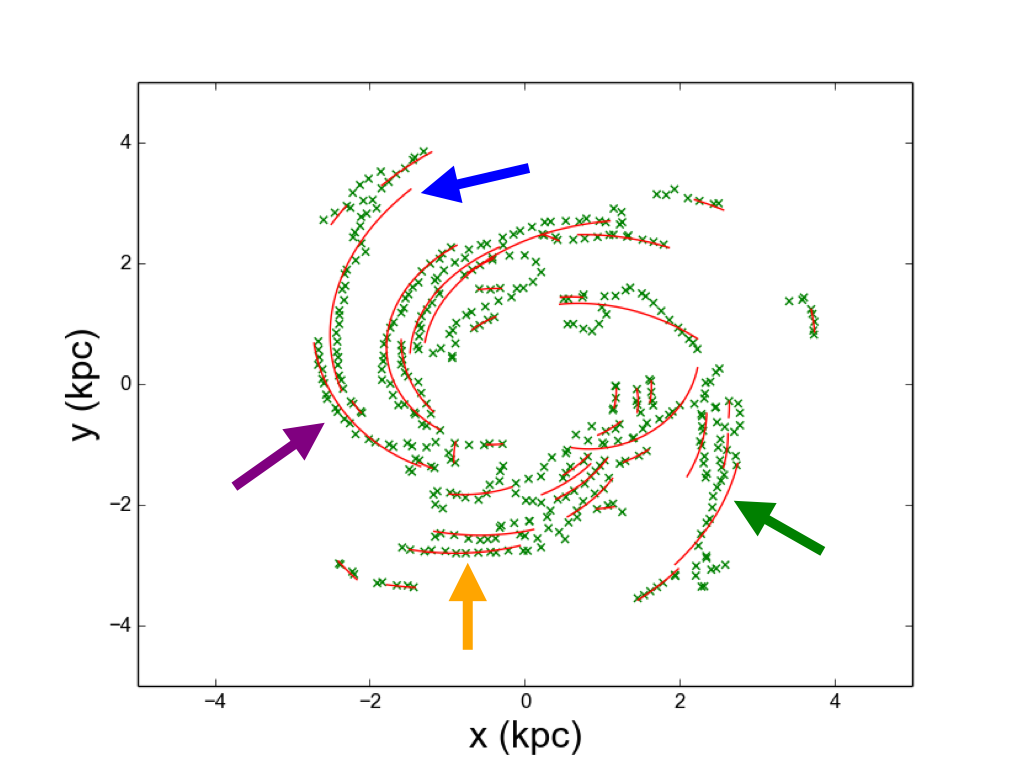} \\
\end{array}$
\end{center}
\caption{The evolution of a galactic disc under a live stellar potential.  Top left: the full SPH simulation containing 2m gas particles, and 2m star particles. Top right: the gas particles identified as ``spiral-like'' (filament class).  Bottom left: gas particles identified as moving in a disc configuration (sheet class).  Bottom right: fits to the spiral structure. Arrows indicate four spiral arms selected for further analysis (see Figure \ref{fig:rotcurve}). \label{fig:livegalaxy}}
\end{figure*}

\begin{figure}
\begin{center}
\includegraphics[scale=0.4]{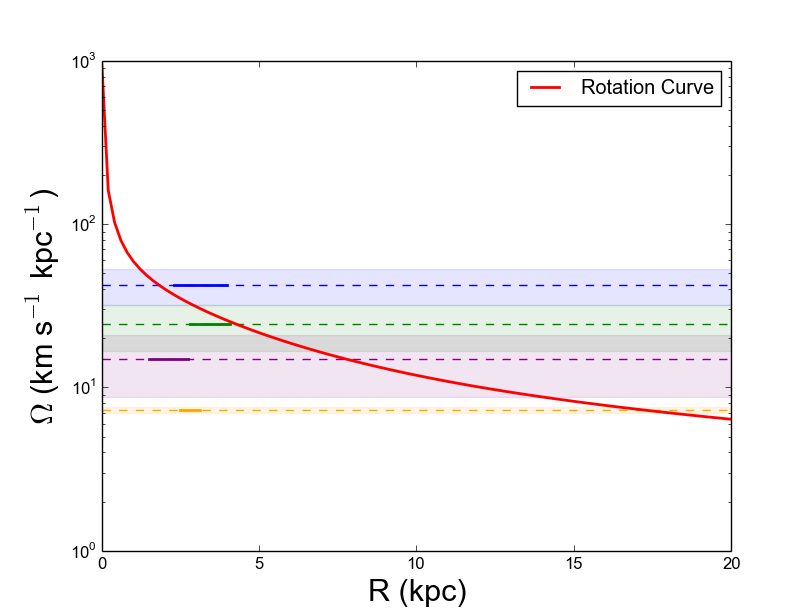}
\end{center}
\caption{The rotation curve of the galactic disc under a live stellar potential, with the pattern speeds of four spiral waves marked by horizontal lines.  Regions with full lines indicate the radial extent of the spiral.  Dashed lines are added to illustrate the corotation radius, denoted by where the horizontal lines meet the rotation curve.  Shaded regions indicate the $1-\sigma$ uncertainties in the pattern speed.  The colours correspond with the arrow indicators in the bottom right panel of Figure \ref{fig:livegalaxy}. \label{fig:rotcurve}}
\end{figure}

\begin{table}
	\centering
	\caption{Galaxy Model Parameters Representative of M33.  The parameters are: $M_d$ - the total baryonic mass of the galactic disc; $R_d, z_d$ - the scale length and scale height of the stellar disc; $f_g$ - the gas fraction of the disc; $f_\star$ - the stellar fraction of the disc; $Q$ - the Toomre parameter; $m$ - expected number of spiral modes (at given radius); $\eta_b$ - the Efstathiou parameter for bar instability, which in this case is less than the critical value; $T$ - the orbital period of the disc at a given radius; $M_b$ - the mass of the bulge; $R_b$ - the scale length of the bulge, $M_h$ - the mass of the dark matter halo; $R_h$ - the scale length of the halo; $c$ - the halo concentration parameter.}
	\begin{center}
		\begin{tabular}{lc}
			\hline
			Disc 					& 		\\
			\hline
			$M_d (10^9\,M_\odot) $		& 9.0		\\
			$R_d (\mathrm{kpc})$		& 2.5		\\
			$z_d (\mathrm{kpc})$		& 0.2		\\
			$f_g$				& 0.15	\\
			$f_\star$				& 0.85	\\
			$Q(1.5 R_d)$				& 1.5		\\
			$m(R=7\mathrm{kpc})$	& 5.6		\\
			$\eta_b$				& 0.9		\\
			$T(R = 2R_d) (\mathrm{Myr})$		& 284	\\
			\hline
			Bulge				&		\\
			\hline
			$M_b (10^8\,M_\odot)$		& 3.0		\\
			$R_b (\mathrm{kpc})$		& 0.4		\\
			\hline
			Halo					&		\\
			\hline
			$M_h (10^{11}\,M_\odot)$	& 5.7		\\
			$R_h (\mathrm{kpc})$		& 33.8	\\
			$c$					& 4.0		\\
			\hline
		\end{tabular}
	\label{table:gal-params}
	\end{center}
\end{table}

We also develop an $N$-body version of the model galaxy, which consists of a live stellar disc, a bulge and a gaseous component. The dark matter halo is represented by a static potential in order to focus the computational efforts in the disc dynamics. The stellar disc is represented by an exponential-isothermal density profile:
\begin{equation} 
	\rho_d(R,z) = \frac{M_d}{4 \pi R_d^2 z_d} \exp\left(-\frac{R}{R_d} \right) \mathrm{sech}^2 \left( \frac{z}{z_d} \right)\,,
\end{equation}
where $R_d$ and $z_d$ are the radial and vertical scale lengths respectively. 

The stellar bulge follows a \citep{Hernquist1990} profile given by:
\begin{equation}
	\rho_b(R) = \frac{M_b }{2 \pi R_b^3} \frac{1}{(R/R_b) (1 + R/R_b)^3}\,,
\end{equation}
where $M_b$ is the mass of the bulge and $R_b$ is the scale radius.

For the dark matter halo, a \citep{NFW1997} profile is used, which has a density profile of the form:
\begin{equation}
	\rho_h(R) = \frac{\rho_0}{(R/R_h) (1 + R/R_h)^2}\,,
\end{equation}
where $R_h$ is the scale radius and $\rho_0$ is a central density parameter. The physical parameters chosen for this model are shown in Table \ref{table:gal-params}, which are based on results from observed data and works modelling the rotation curve of M33 (e.~g. \citealt{ReganVogel1994,CorbelliSalucci2000,Seigar2011,HagueWilkinson2015,Kametal2015})

The initial conditions are obtained by using the method of \citet{McMillanDehnen2007}, which is publicly available through the code \emph{mkgalaxy} \footnote{The \emph{mkgalaxy} code can be obtained from the NEMO Stellar Dynamics Toolbox package \citep{Teuben1995} (https://bima.astro.umd.edu/nemo/).}. It generates a self-consistent model of a galaxy composed by a halo, disc, and bulge. It has the advantage of creating a stellar disc with initial velocities sampled from a distribution function representative of a disc rather than assuming a local Maxwellian distribution in velocity space \citep{Dehnen1999b}. The disc particles have a velocity distribution consistent with the rotation curve of the galaxy. However, this code only produces a collisionless model of the galaxy and the gas initialisation has to be treated separately.

To build the full galaxy model, a disc component containing the total number of particles and the total combined stellar and gas mass is first generated. Then, to initialise the gaseous component, the original disc is divided into the two groups of particles and are assigned the following particle masses: $m_g = M_g/N_g, m_\star=M_\star/N_\star$ for the gas and stellar components, respectively. At this point, the model has a gas and a stellar component, with the positions and velocities obtained from the initial conditions generator. This is not an equilibrium condition for the gas itself, so the system has to be evolved for some time in order to allow it to settle in the galaxy. This scheme has been previously tested in \citet{RamonFox2014} and a similar method also in \citet{Dobbs2010}.

Figure \ref{fig:livegalaxy} shows the full simulation, its arm and inter-arm components, and fits to the spiral structure.  The flocculent spiral structures are clearly identified by classification using the velocity shear tensor ($E=2$), but the spine fitting algorithm only outlines the larger arms with any clarity.  

In Table \ref{table:gal-spirals} we presented calculated pitch angles and pattern speeds for the longest wave in the upper left and lower right quadrants the galaxy, as well as two other shorter waves in the lower left quadrant (identified by arrows in the bottom right panel of Figure \ref{fig:livegalaxy}).  The long waves yield corotation radii of approximately 1.8 and 4.2 kpc respectively (Figure \ref{fig:rotcurve}).  These values bracket the scale length of the stellar disc ($R_d = 2.5 \,\rm{kpc}$), and denote the region where the rotation curves of both the stellar and gaseous components begin to flatten, and the stellar bulge and gaseous disc enter corotation.

\begin{table}
	\centering
	\caption{Fits to the four spiral arms selected from the galaxy simulation with a live stellar potential.  The arms 1, 2, 3 and 4 correspond to the blue, green, orange and purple arrows in Figure \ref{fig:livegalaxy} (and similarly for the lines in Figure \ref{fig:rotcurve}).  We compute uncertainties for $R_{\rm corot}$ by computing its value at the upper and lower $1\sigma$ limits for $\Omega_p$.}
	\begin{center}
		\begin{tabular}{ccccc}
			\hline
		Arm &  $\phi$ ($^{\circ}$) & $\Omega_p$ (km s$^{-1}$ kpc$^{-1}$) & $R_{\rm corot}$ (kpc) \\ 	
		\hline		
			1 & 17.59 $\pm$ 0.87 & 42.57 $\pm$ 10.59 & 1.82$^{+1.01}_{-0.61}$ \\
			2 & 16.81 $\pm$ 0.85 & 24.34 $\pm$ 7.54 & 4.25$^{+2.62}_{-1.41}$ \\
			3 & 13.01 $\pm$ 0.46 & 7.27 $\pm$ 0.28 & 17.17$^{+0.81}_{-0.61}$ \\
			4 & 16.88 $\pm$ 1.52 & 14.86 $\pm$ 6.14 & 7.88$^{+6.05}_{-2.62}$  \\
			\hline
		\end{tabular}
	\label{table:gal-spirals}
	\end{center}
\end{table}

The smaller spiral waves reside inside this radius, and appear to have much longer corotation radii (approximately 7 and 17 kpc respectively).  These appear to be less robust, transient features.


\section{Discussion}\label{sec:discussion}

By identifying simulation regions that are ``spiral-like'', we can use the complement of these regions to identify interarm regions.  This is a valuable distinction for studies of star formation in spiral galaxies.  We can use this technique to trace the passage of a spiral shock through gas, recording the density, temperature and chemical evolution of the gas as it forms molecular clouds, and eventually stars.  For example, as gas transitions from the warm diffuse phase into the dense cool phase, we will be able to time this transition relative to when the gas entered and left a spiral structure.

This will be of great use in investigating the deviation of low-surface density clouds from the Kennicutt-Schmidt star formation relation \citep{Bonnell2013}, and the relationship in general between spiral structure and molecular clouds \citep{Duarte-Cabral2016}.  Spiral structure also drives amplification and reversal of the galactic magnetic field, particularly at large velocity jumps over the spiral shock and at corotation \citep{Dobbs2016}.  This process (of relevance to the boundness of molecular clouds) can be studied in finer detail if the fluid elements driving these magnetic amplifications and reversals can be more rigorously identified.

Being able to separate the interarm component from the arm component has its own benefits.  When computing ``unperturbed'' disc properties, it is common to simply take azimuthal averages of the disc, and assuming the perturbed component is insignificant compared to the wider disc.  If the spiral perturbation amplitude is large, this assumption fails.  Being able to compute averages in the knowledge that the spiral structure has been extracted will allow a more accurate computation of the spiral perturbation itself ($\Delta \Sigma/\Sigma$), as well as a decontaminated study of the unperturbed disc.  \citet{Foyle2011} note that star formation rate tracers can be prominent in the interarm regions of local spiral galaxies -- effective characterisation of interarm regions in simulations are needed to investigate this phenomenon, and link it to the triggering of star formation by spiral arms, as well as \emph{in situ} star formation processes.

Our spiral identification technique can also be applied to the stellar component (provided that the velocity shear tensor can be appropriately  calculated for individual star particles).  This will allow us to measure offsets between stellar and gaseous spiral structure (cf \citealt{Pettitt2016}), as well as track the motion of stars relative to both the gaseous and stellar arms.  This has extra relevance to attempts to elucidate the Milky Way's spiral structure.  For example, using molecular clouds to trace spiral arms has demonstrable systematic offsets due to use of the kinematic distance \citep{Ramon-Fox2017}.

The feeding of active galactic nuclei (AGN) begins with the outward transport of angular momentum at large scales, allowing matter to flow into the inner 2-3 kpc (where stellar bars can take over).  Lagrangian methods like SPH will be able to trace a fluid element's journey from large distances towards the nucleus, recording its encounters with spiral structure along the way.

Spiral classification has important benefits for the study of self-gravitating protostellar discs, and discs which are perturbed by internal and external companions.  In the first instance, full characterisation of a spiral arm can identify whether it is being self-generated by the disc or is due to a companion \citep{Meru2017}.  Strong spiral arms in protostellar discs are typically induced by self-gravity, but non-axisymmetric structure can also be induced by the magneto-rotational instability (e.g. \citealt{Nelson2005}).  Spiral classification will provide important discriminants between the two.

Spiral arms in protostellar discs drive chemical evolution \citep{Evans2015} and potentially grain growth \citep{Rice2004,Booth2016}.  Timing the perturbation of gas properties (and dust particles) to its passage through spiral arm passage will be critical to understanding their effectiveness.

Spiral classification is also valuable for the study of fragmentation in self-gravitating discs.  The interaction of fragments with spiral arms has a significant effect on their survival rate \citep{Hall2017} and their chemical inventory \citep{Ilee2017}.  Being able to observe how material is transferred between the spiral arm and the fragment will provide important insights into how fragments accrete, and how their orbital elements, physical properties and chemistry is governed by interactions with spiral structure.

Spirals driven by companions in protostellar discs offer key diagnostics of the disc and the companion, as mentioned in the Introduction.  Automatic characterisation of spirals allows simulators to run increasingly large banks of simulations over a wider parameter space.  This opens up the possibility of potentially identifying new diagnostics from spiral arm data, as well as improved fitting of observations from grids of model runs.

Throughout this paper, we have focused on classification via a single tensor, and the choice of tensor has depended largely on the physics at play.  In the majority of cases, where the spiral perturbation amplitude is relatively weak, spiral detection proceeds best via the velocity shear tensor (and identifying $E=2$ particles).  In the limit where surface density perturbations are large, tidal tensor analysis is usually preferred. Our example of the self-gravitating protostellar disc is relatively extreme, in that we find best results for $E=3$ particles.  Our choices have reflected some knowledge of the physics of the system - the converse is equally true, that analysing using different tensors yields insight into the governing physics.

In \citet{Forgan2016}, we showed the benefits of classifying a system using multiple tensors, and then correlating the data.  In an example showing the classification of structure around a supernova explosion.  The velocity shear tensor identifies regions at the inner edge of the cavity driven by the supernova; the tidal tensor identifies regions being swept into collapsing structures; correlating data identifies regions under collapse as a direct result of the supernova explosion.  Spiral detection under multiple tensors can yield important information about the relative amplitudes of individual arms.

Finally, it is worth noting that our algorithm is made of two largely separate components: tensor classification and spiral spine identification.  The latter component can be swapped for another algorithm depending on the user's aims.  Our spine identification approach works best on the most dense regions of the simulation, and is therefore not well suited to the outer regions (as can be seen by its failure to find some structures visible ``by eye'').  Ideally, a spine identification algorithm should use the tensor adopted for classification.  For example, arms detected using the velocity shear tensor should be detected using percentiles of some measure of velocity shear (say the maximum eigenvalue) rather than percentiles of density.  We leave this as a route to follow in future work.

\section{Conclusions}\label{sec:conclusions}

\noindent We have demonstrated analysis techniques that isolate and fit individual spiral arms in astrophysical disc simulations.  For simulations of both protostellar and galactic discs, we are able to fit logarithmic spirals to each arm, and compute pitch angles.  If multiple snapshots are available, we are also able to compute the pattern speed of spiral density waves, and identify where the wave corotates with the bulk flow.

A key innovation of these techniques is the ability to identify when fluid elements are inside a spiral structure, or in the interarm region.

Our techniques rely purely on computing derivatives of the flow at a spatial location, and the subsequent application of friends-of-friends algorithms.  Therefore, our approach is applicable to data from any finite spatial element hydrodynamic solver, such as grid-based solvers using adaptive mesh refinement, such as ENZO \citep{Bryan2014}, Voronoi mesh solvers such as AREPO \citep{Springel2010}, or indeed meshless hydrodynamic systems such as GIZMO \citep{Hopkins2015}.  We expect our methods will prove extremely useful to simulators studying the role of spirals in the evolution of disc structures throughout the Universe.

\section*{Acknowledgments}

DHF, FGR-F and IAB gratefully acknowledge support from the ECOGAL project, grant agreement 291227, funded by the European Research Council under ERC-2011-ADG.  This  research  has  made  use  of  NASA's  Astrophysics  Data  System Bibliographic  Services.  This work relied on the compute resources of the St Andrews MHD Cluster, and the DiRAC Complexity system, operated by the University of Leicester IT Services, which forms part of the STFC DiRAC HPC facility (\url{www.dirac.ac.uk}).  Surface density plots were created using \texttt{SPLASH} \citep{SPLASH}.  Corner plots were produced using the \texttt{corner.py} module \citep{corner}.  Our code is published on Github at \url{https://github.com/dh4gan/tache}.  The authors warmly thank the anonymous reviewer for their careful reading and astute suggestions.

\bibliographystyle{mnras} 
\bibliography{sph_spiral}

\begin{thebibliography}{}
\makeatletter
\relax
\def\mn@urlcharsother{\let\do\@makeother \do\$\do\&\do\#\do\^\do\_\do\%\do\~}
\def\mn@doi{\begingroup\mn@urlcharsother \@ifnextchar [ {\mn@doi@}
  {\mn@doi@[]}}
\def\mn@doi@[#1]#2{\def\@tempa{#1}\ifx\@tempa\@empty \href
  {http://dx.doi.org/#2} {doi:#2}\else \href {http://dx.doi.org/#2} {#1}\fi
  \endgroup}
\def\mn@eprint#1#2{\mn@eprint@#1:#2::\@nil}
\def\mn@eprint@arXiv#1{\href {http://arxiv.org/abs/#1} {{\tt arXiv:#1}}}
\def\mn@eprint@dblp#1{\href {http://dblp.uni-trier.de/rec/bibtex/#1.xml}
  {dblp:#1}}
\def\mn@eprint@#1:#2:#3:#4\@nil{\def\@tempa {#1}\def\@tempb {#2}\def\@tempc
  {#3}\ifx \@tempc \@empty \let \@tempc \@tempb \let \@tempb \@tempa \fi \ifx
  \@tempb \@empty \def\@tempb {arXiv}\fi \@ifundefined
  {mn@eprint@\@tempb}{\@tempb:\@tempc}{\expandafter \expandafter \csname
  mn@eprint@\@tempb\endcsname \expandafter{\@tempc}}}

\bibitem[\protect\citeauthoryear{Alexander \& Hickox}{Alexander \&
  Hickox}{2012}]{Alexander2012}
Alexander D.,  Hickox R.,  2012, \mn@doi [New Astronomy Reviews]
  {10.1016/j.newar.2011.11.003}, 56, 93

\bibitem[\protect\citeauthoryear{Artymowicz \& Lubow}{Artymowicz \&
  Lubow}{1994}]{Artymowicz1994}
Artymowicz P.,  Lubow S.~H.,  1994, \mn@doi [ApJ] {10.1086/173679}, 421, 651

\bibitem[\protect\citeauthoryear{Bate, Bonnell  \& Bromm}{Bate
  et~al.}{2003}]{Bate2002}
Bate M.~R.,  Bonnell I.~A.,   Bromm V.,  2003, \mn@doi [MNRAS]
  {10.1046/j.1365-8711.2003.06210.x}, 339, 577

\bibitem[\protect\citeauthoryear{Benisty et~al.,}{Benisty
  et~al.}{2015}]{Benisty2015}
Benisty M.,  et~al., 2015, \mn@doi [Astronomy {\&} Astrophysics]
  {10.1051/0004-6361/201526011}, 578, L6

\bibitem[\protect\citeauthoryear{Binney \& Tremaine}{Binney \&
  Tremaine}{2008}]{BandT2008}
Binney J.,  Tremaine S.,  2008, {Galactic Dynamics}.
Princeton University Press

\bibitem[\protect\citeauthoryear{Bonnell \& Bate}{Bonnell \&
  Bate}{1994}]{Bonnell1994}
Bonnell I.~A.,  Bate M.~R.,  1994, MNRAS, 271, 999

\bibitem[\protect\citeauthoryear{Bonnell, Dobbs, Robitaille  \&
  Pringle}{Bonnell et~al.}{2006}]{Bonnell2006}
Bonnell I.~A.,  Dobbs C.~L.,  Robitaille T.~P.,   Pringle J.~E.,  2006, \mn@doi
  [MNRAS] {10.1111/j.1365-2966.2005.09657.x}, 365, 37

\bibitem[\protect\citeauthoryear{Bonnell, Dobbs  \& Smith}{Bonnell
  et~al.}{2013}]{Bonnell2013}
Bonnell I.~A.,  Dobbs C.~L.,   Smith R.~J.,  2013, MNRAS, 430, 1790

\bibitem[\protect\citeauthoryear{Booth \& Clarke}{Booth \&
  Clarke}{2016}]{Booth2016}
Booth R.~A.,  Clarke C.~J.,  2016, \mn@doi [MNRAS] {10.1093/mnras/stw488}, 458,
  2676

\bibitem[\protect\citeauthoryear{Bryan et~al.,}{Bryan et~al.}{2014}]{Bryan2014}
Bryan G.~L.,  et~al., 2014, \mn@doi [ApJ] {10.1088/0067-0049/211/2/19}, 211, 19

\bibitem[\protect\citeauthoryear{Choi, Dalcanton, Williams, Weisz, Skillman,
  Fouesneau  \& Dolphin}{Choi et~al.}{2015}]{Choi2015}
Choi Y.,  Dalcanton J.~J.,  Williams B.~F.,  Weisz D.~R.,  Skillman E.~D.,
  Fouesneau M.,   Dolphin A.~E.,  2015, \mn@doi [The Astrophysical Journal]
  {10.1088/0004-637X/810/1/9}, 810, 9

\bibitem[\protect\citeauthoryear{Clarke \& Lodato}{Clarke \&
  Lodato}{2009}]{Clarke2009}
Clarke C.~J.,  Lodato G.,  2009, \mn@doi [MNRAS]
  {10.1111/j.1745-3933.2009.00695.x}, 398, L6

\bibitem[\protect\citeauthoryear{Corbelli \& Salucci}{Corbelli \&
  Salucci}{2000}]{CorbelliSalucci2000}
Corbelli E.,  Salucci P.,  2000, MNRAS, 311, 441

\bibitem[\protect\citeauthoryear{Cossins, Lodato  \& Clarke}{Cossins
  et~al.}{2009}]{Cossins2008}
Cossins P.,  Lodato G.,   Clarke C.~J.,  2009, \mn@doi [MNRAS]
  {10.1111/j.1365-2966.2008.14275.x}, 393, 1157

\bibitem[\protect\citeauthoryear{Cox \& G{\'{o}}mez}{Cox \&
  G{\'{o}}mez}{2002}]{CoxandGomez2002}
Cox D.,  G{\'{o}}mez G.,  2002, ApJS, 142, 261

\bibitem[\protect\citeauthoryear{Davis et~al.,}{Davis et~al.}{2015}]{Davis2015}
Davis B.~L.,  et~al., 2015, \mn@doi [ApJ] {10.1088/2041-8205/802/1/L13}, 802,
  L13

\bibitem[\protect\citeauthoryear{Dehnen}{Dehnen}{1999}]{Dehnen1999b}
Dehnen W.,  1999, AJ, 118, 1201

\bibitem[\protect\citeauthoryear{Dobbs \& Baba}{Dobbs \&
  Baba}{2014}]{Dobbs2014b}
Dobbs C.,  Baba J.,  2014, \mn@doi [Publications of the Astronomical Society of
  Australia] {10.1017/pasa.2014.31}, 31, e035

\bibitem[\protect\citeauthoryear{Dobbs \& Bonnell}{Dobbs \&
  Bonnell}{2007}]{Dobbs2007}
Dobbs C.~L.,  Bonnell I.~A.,  2007, \mn@doi [MNRAS]
  {10.1111/j.1365-2966.2006.11227.x}, 374, 1115

\bibitem[\protect\citeauthoryear{Dobbs, Theis, Pringle  \& Bate}{Dobbs
  et~al.}{2010}]{Dobbs2010}
Dobbs C.~L.,  Theis C.,  Pringle J.~E.,   Bate M.~R.,  2010, \mn@doi [MNRAS]
  {10.1111/j.1365-2966.2009.16161.x}, 403, 625

\bibitem[\protect\citeauthoryear{Dobbs, Price, Pettitt, Bate  \& Tricco}{Dobbs
  et~al.}{2016}]{Dobbs2016}
Dobbs C.~L.,  Price D.~J.,  Pettitt A.~R.,  Bate M.~R.,   Tricco T.~S.,  2016,
  \mn@doi [MNRAS] {10.1093/mnras/stw1625}, 461, 4482

\bibitem[\protect\citeauthoryear{Dong, Hall, Rice  \& Chiang}{Dong
  et~al.}{2015}]{Dong2015}
Dong R.,  Hall C.,  Rice K.,   Chiang E.,  2015, \mn@doi [The Astrophysical
  Journal] {10.1088/2041-8205/812/2/L32}, 812, L32

\bibitem[\protect\citeauthoryear{Duarte-Cabral \& Dobbs}{Duarte-Cabral \&
  Dobbs}{2016}]{Duarte-Cabral2016}
Duarte-Cabral A.,  Dobbs C.~L.,  2016, \mn@doi [MNRAS] {10.1093/mnras/stw469},
  458, 3667

\bibitem[\protect\citeauthoryear{Egusa, Sofue  \& Nakanishi}{Egusa
  et~al.}{2004}]{Egusa2004}
Egusa F.,  Sofue Y.,   Nakanishi H.,  2004, \mn@doi [Publications of the
  Astronomical Society of Japan] {10.1093/pasj/56.6.L45}, 56, L45

\bibitem[\protect\citeauthoryear{Egusa, Kohno, Sofue, Nakanishi  \&
  Komugi}{Egusa et~al.}{2009}]{Egusa2009}
Egusa F.,  Kohno K.,  Sofue Y.,  Nakanishi H.,   Komugi S.,  2009, \mn@doi [The
  Astrophysical Journal] {10.1088/0004-637X/697/2/1870}, 697, 1870

\bibitem[\protect\citeauthoryear{Elmegreen}{Elmegreen}{1990}]{Elmegreen1990}
Elmegreen B.~G.,  1990, \mn@doi [Annals of the New York Academy of Sciences]
  {10.1111/j.1749-6632.1990.tb27410.x}, 596, 40

\bibitem[\protect\citeauthoryear{Evans, Ilee, Boley, Caselli, Durisen,
  Hartquist  \& Rawlings}{Evans et~al.}{2015}]{Evans2015}
Evans M.~G.,  Ilee J.~D.,  Boley A.~C.,  Caselli P.,  Durisen R.~H.,  Hartquist
  T.~W.,   Rawlings J. M.~C.,  2015, \mn@doi [MNRAS] {10.1093/mnras/stv1698},
  453, 1147

\bibitem[\protect\citeauthoryear{Foreman-Mackey}{Foreman-Mackey}{2016}]{corner}
Foreman-Mackey D.,  2016, \mn@doi [The Journal of Open Source Software]
  {10.21105/joss.00024}, 24

\bibitem[\protect\citeauthoryear{Forero-Romero, Hoffman, Gottl{\"{o}}ber,
  Klypin  \& Yepes}{Forero-Romero et~al.}{2009}]{Forero-Romero2009}
Forero-Romero J.~E.,  Hoffman Y.,  Gottl{\"{o}}ber S.,  Klypin a.,   Yepes G.,
  2009, \mn@doi [MNRAS] {10.1111/j.1365-2966.2009.14885.x}, 396, 1815

\bibitem[\protect\citeauthoryear{Forgan \& Rice}{Forgan \&
  Rice}{2011}]{Forgan2011a}
Forgan D.,  Rice K.,  2011, \mn@doi [MNRAS] {10.1111/j.1365-2966.2011.19380.x},
  417, 1928

\bibitem[\protect\citeauthoryear{Forgan \& Rice}{Forgan \&
  Rice}{2013}]{TD_synthesis}
Forgan D.,  Rice K.,  2013, MNRAS, 432, 3168

\bibitem[\protect\citeauthoryear{Forgan, Rice, Stamatellos  \&
  Whitworth}{Forgan et~al.}{2009}]{intro_hybrid}
Forgan D.~H.,  Rice K.,  Stamatellos D.,   Whitworth A.~P.,  2009, \mn@doi
  [MNRAS] {10.1111/j.1365-2966.2008.14373.x}, 394, 882

\bibitem[\protect\citeauthoryear{Forgan, Rice, Cossins  \& Lodato}{Forgan
  et~al.}{2011}]{Forgan2011}
Forgan D.,  Rice K.,  Cossins P.,   Lodato G.,  2011, \mn@doi [MNRAS]
  {10.1111/j.1365-2966.2010.17500.x}, 410, 994

\bibitem[\protect\citeauthoryear{Forgan, Parker  \& Rice}{Forgan
  et~al.}{2015}]{TD_dynamics}
Forgan D.,  Parker R.~J.,   Rice K.,  2015, MNRAS, 447, 836

\bibitem[\protect\citeauthoryear{Forgan, Bonnell, Lucas  \& Rice}{Forgan
  et~al.}{2016a}]{Forgan2016}
Forgan D.,  Bonnell I.,  Lucas W.,   Rice K.,  2016a, MNRAS, 457, 2501

\bibitem[\protect\citeauthoryear{Forgan, Ilee, Cyganowski, Brogan  \&
  Hunter}{Forgan et~al.}{2016b}]{Forgan2016e}
Forgan D.~H.,  Ilee J.~D.,  Cyganowski C.~J.,  Brogan C.~L.,   Hunter T.~R.,
  2016b, MNRAS, 463, 957

\bibitem[\protect\citeauthoryear{Foyle, Rix, Walter  \& Leroy}{Foyle
  et~al.}{2010}]{Foyle2010}
Foyle K.,  Rix H.-W.,  Walter F.,   Leroy A.~K.,  2010, \mn@doi [The
  Astrophysical Journal] {10.1088/0004-637X/725/1/534}, 725, 534

\bibitem[\protect\citeauthoryear{Foyle, Rix, Dobbs, Leroy  \& Walter}{Foyle
  et~al.}{2011}]{Foyle2011}
Foyle K.,  Rix H.-W.,  Dobbs C.~L.,  Leroy A.~K.,   Walter F.,  2011, \mn@doi
  [The Astrophysical Journal] {10.1088/0004-637X/735/2/101}, 735, 101

\bibitem[\protect\citeauthoryear{Grand, Kawata  \& Cropper}{Grand
  et~al.}{2012}]{Grand2012}
Grand R. J.~J.,  Kawata D.,   Cropper M.,  2012, \mn@doi [MNRAS]
  {10.1111/j.1365-2966.2012.21733.x}, 426, 167

\bibitem[\protect\citeauthoryear{Hague \& Wilkinson}{Hague \&
  Wilkinson}{2015}]{HagueWilkinson2015}
Hague P.,  Wilkinson M.,  2015, ApJ, 800, id

\bibitem[\protect\citeauthoryear{Hahn, Porciani, Carollo  \& Dekel}{Hahn
  et~al.}{2007}]{Hahn2007}
Hahn O.,  Porciani C.,  Carollo C.~M.,   Dekel A.,  2007, \mn@doi [MNRAS]
  {10.1111/j.1365-2966.2006.11318.x}, 375, 489

\bibitem[\protect\citeauthoryear{Hall, Forgan  \& Rice}{Hall
  et~al.}{2017}]{Hall2017}
Hall C.,  Forgan D.,   Rice K.,  2017, \mn@doi [MNRAS] {10.1093/mnras/stx1244},
  470, 2517

\bibitem[\protect\citeauthoryear{Hernquist}{Hernquist}{1990}]{Hernquist1990}
Hernquist L.,  1990, ApJ, 356, 359

\bibitem[\protect\citeauthoryear{Heyer \& Dame}{Heyer \&
  Dame}{2015}]{Heyer2015}
Heyer M.,  Dame T.,  2015, \mn@doi [Annual Review of Astronomy and
  Astrophysics] {10.1146/annurev-astro-082214-122324}, 53, 583

\bibitem[\protect\citeauthoryear{Holmberg}{Holmberg}{1941}]{Holmberg1941}
Holmberg E.,  1941, \mn@doi [ApJ] {10.1086/144344}, 94, 385

\bibitem[\protect\citeauthoryear{Hopkins}{Hopkins}{2015}]{Hopkins2015}
Hopkins P.~F.,  2015, \mn@doi [MNRAS] {10.1093/mnras/stv195}, 450, 53

\bibitem[\protect\citeauthoryear{Hubble}{Hubble}{1926}]{Hubble1926}
Hubble E.~P.,  1926, \mn@doi [The Astrophysical Journal] {10.1086/143018}, 64,
  321

\bibitem[\protect\citeauthoryear{Ilee, Boley, Caselli, Durisen, Hartquist  \&
  Rawlings}{Ilee et~al.}{2011}]{Ilee2011}
Ilee J.~D.,  Boley A.~C.,  Caselli P.,  Durisen R.~H.,  Hartquist T.~W.,
  Rawlings J. M.~C.,  2011, \mn@doi [MNRAS] {10.1111/j.1365-2966.2011.19455.x},
  417, 2950

\bibitem[\protect\citeauthoryear{Ilee et~al.,}{Ilee et~al.}{2017}]{Ilee2017}
Ilee J.~D.,  et~al., 2017, MNRAS, 472, 189

\bibitem[\protect\citeauthoryear{{Kennicutt, R. C.}}{{Kennicutt, R.
  C.}}{1981}]{Kennicutt1981}
{Kennicutt, R. C.} J.,  1981, \mn@doi [The Astronomical Journal]
  {10.1086/113064}, 86, 1847

\bibitem[\protect\citeauthoryear{Laughlin \& Rozyczka}{Laughlin \&
  Rozyczka}{1996}]{Laughlin1996}
Laughlin G.,  Rozyczka M.,  1996, \mn@doi [ApJ] {10.1086/176648}, 456, 279

\bibitem[\protect\citeauthoryear{Lin \& Shu}{Lin \& Shu}{1964}]{Lin1964}
Lin C.~C.,  Shu F.~H.,  1964, \mn@doi [ApJ] {10.1086/147955}, 140, 646

\bibitem[\protect\citeauthoryear{Lintott et~al.,}{Lintott
  et~al.}{2011}]{Lintott2011}
Lintott C.,  et~al., 2011, \mn@doi [Monthly Notices of the Royal Astronomical
  Society] {10.1111/j.1365-2966.2010.17432.x}, 410, 166

\bibitem[\protect\citeauthoryear{Lodato \& Price}{Lodato \&
  Price}{2010}]{Lodato2010}
Lodato G.,  Price D.~J.,  2010, \mn@doi [MNRAS]
  {10.1111/j.1365-2966.2010.16526.x}, 405, 1212

\bibitem[\protect\citeauthoryear{Lodato \& Rice}{Lodato \&
  Rice}{2004}]{Lodato_and_Rice_04}
Lodato G.,  Rice W. K.~M.,  2004, \mn@doi [MNRAS]
  {10.1111/j.1365-2966.2004.07811.x}, 351, 630

\bibitem[\protect\citeauthoryear{Lodato \& Rice}{Lodato \&
  Rice}{2005}]{Lodato2005}
Lodato G.,  Rice W. K.~M.,  2005, \mn@doi [MNRAS]
  {10.1111/j.1365-2966.2005.08875.x}, 358, 1489

\bibitem[\protect\citeauthoryear{Mata-Chavez, Gomez  \& Puerari}{Mata-Chavez
  et~al.}{2014}]{Mata-Chavez2014}
Mata-Chavez M.~D.,  Gomez G.~C.,   Puerari I.,  2014, \mn@doi [MNRAS]
  {10.1093/mnras/stu1672}, 444, 3756

\bibitem[\protect\citeauthoryear{McMillan \& Dehnen}{McMillan \&
  Dehnen}{2007}]{McMillanDehnen2007}
McMillan P.~J.,  Dehnen W.,  2007, MNRAS, 378, 541

\bibitem[\protect\citeauthoryear{Meru, Juh{\'{a}}sz, Ilee, Clarke, Rosotti  \&
  Booth}{Meru et~al.}{2017}]{Meru2017}
Meru F.,  Juh{\'{a}}sz A.,  Ilee J.~D.,  Clarke C.~J.,  Rosotti G.~P.,   Booth
  R.~A.,  2017, \mn@doi [ApJ] {10.3847/2041-8213/aa6837}, 839, L24

\bibitem[\protect\citeauthoryear{Monaghan}{Monaghan}{1992}]{Monaghan_92}
Monaghan J.~J.,  1992, \mn@doi [ARA{\&}A]
  {10.1146/annurev.aa.30.090192.002551}, 30, 543

\bibitem[\protect\citeauthoryear{Monaghan}{Monaghan}{2005}]{Monaghan_05}
Monaghan J.~J.,  2005, \mn@doi [Reports on Progress in Physics]
  {10.1088/0034-4885/68/8/R01}, 68, 1703

\bibitem[\protect\citeauthoryear{Murray}{Murray}{1996}]{Murray1996}
Murray J.~R.,  1996, MNRAS, 279, 402

\bibitem[\protect\citeauthoryear{Muto et~al.,}{Muto et~al.}{2012}]{Muto2012}
Muto T.,  et~al., 2012, \mn@doi [The Astrophysical Journal]
  {10.1088/2041-8205/748/2/L22}, 748, L22

\bibitem[\protect\citeauthoryear{Navarro, Frenk  \& White}{Navarro
  et~al.}{1997}]{NFW1997}
Navarro J.,  Frenk C.,   White S.,  1997, ApJ, 490, 493

\bibitem[\protect\citeauthoryear{Nelson}{Nelson}{2005}]{Nelson2005}
Nelson R.~P.,  2005, \mn@doi [Astronomy {\&} Astrophysics]
  {10.1051/0004-6361:20042605}, 443, 1067

\bibitem[\protect\citeauthoryear{Papaloizou \& Savonije}{Papaloizou \&
  Savonije}{1991}]{Papaloizou1991}
Papaloizou J.~C.,  Savonije G.~J.,  1991, \mn@doi [MNRAS]
  {10.1093/mnras/248.3.353}, 248, 353

\bibitem[\protect\citeauthoryear{P{\'{e}}rez et~al.,}{P{\'{e}}rez
  et~al.}{2016}]{Perez2016}
P{\'{e}}rez L.~M.,  et~al., 2016, Science, 353

\bibitem[\protect\citeauthoryear{Pettitt, Tasker  \& Wadsley}{Pettitt
  et~al.}{2016}]{Pettitt2016}
Pettitt A.~R.,  Tasker E.~J.,   Wadsley J.~W.,  2016, \mn@doi [MNRAS]
  {10.1093/mnras/stw588}, 458, 3990

\bibitem[\protect\citeauthoryear{Pohl, Pinilla, Benisty, Ataiee, Juh{\'{a}}sz,
  Dullemond, {Van Boekel}  \& Henning}{Pohl et~al.}{2015}]{Pohl2015}
Pohl A.,  Pinilla P.,  Benisty M.,  Ataiee S.,  Juh{\'{a}}sz A.,  Dullemond
  C.~P.,  {Van Boekel} R.,   Henning T.,  2015, \mn@doi [MNRAS]
  {10.1093/mnras/stv1746}, 453, 1768

\bibitem[\protect\citeauthoryear{Price}{Price}{2007}]{SPLASH}
Price D.~J.,  2007, \mn@doi [PASA] {10.1071/AS07022}, 24, 159

\bibitem[\protect\citeauthoryear{Price}{Price}{2012}]{Price2012}
Price D.~J.,  2012, \mn@doi [Journal of Computational Physics]
  {10.1016/j.jcp.2010.12.011}, 231, 759

\bibitem[\protect\citeauthoryear{Querejeta et~al.,}{Querejeta
  et~al.}{2016}]{Querejeta2016}
Querejeta M.,  et~al., 2016, \mn@doi [Astronomy {\&} Astrophysics]
  {10.1051/0004-6361/201527536}, 588, A33

\bibitem[\protect\citeauthoryear{Rafikov}{Rafikov}{2002}]{Rafikov2002}
Rafikov R.~R.,  2002, \mn@doi [The Astrophysical Journal] {10.1086/339399},
  569, 997

\bibitem[\protect\citeauthoryear{Ragan, Moore, Eden, Hoare, Elia  \&
  Molinari}{Ragan et~al.}{2016}]{Ragan2016}
Ragan S.~E.,  Moore T. J.~T.,  Eden D.~J.,  Hoare M.~G.,  Elia D.,   Molinari
  S.,  2016, \mn@doi [MNRAS] {10.1093/mnras/stw1870}, 462, 3123

\bibitem[\protect\citeauthoryear{Ram{\'{o}}n-Fox \& Aceves}{Ram{\'{o}}n-Fox \&
  Aceves}{2014}]{RamonFox2014}
Ram{\'{o}}n-Fox F.,  Aceves H.,  2014, in Seigar M.,  Treuthardt P.,  eds,
  Structure and Dynamics of Disk Galaxies. Astronomical Society of the Pacific,
  p.~229

\bibitem[\protect\citeauthoryear{Ram{\'{o}}n-Fox \& Bonnell}{Ram{\'{o}}n-Fox \&
  Bonnell}{2018}]{Ramon-Fox2017}
Ram{\'{o}}n-Fox F.~G.,  Bonnell I.~A.,  2018, \mn@doi [MNRAS]
  {10.1093/mnras/stx2866}, 474, 2028

\bibitem[\protect\citeauthoryear{Regan \& Vogel}{Regan \&
  Vogel}{1994}]{ReganVogel1994}
Regan M.,  Vogel S.,  1994, ApJ, 434, 536

\bibitem[\protect\citeauthoryear{Rice \& Armitage}{Rice \&
  Armitage}{2009}]{Rice_and_Armitage_09}
Rice W. K.~M.,  Armitage P.~J.,  2009, \mn@doi [MNRAS]
  {10.1111/j.1365-2966.2009.14879.x}, 396, 2228

\bibitem[\protect\citeauthoryear{Rice, Lodato, Pringle, Armitage  \&
  Bonnell}{Rice et~al.}{2004}]{Rice2004}
Rice W. K.~M.,  Lodato G.,  Pringle J.~E.,  Armitage P.~J.,   Bonnell I.~A.,
  2004, MNRAS, 355, 543

\bibitem[\protect\citeauthoryear{Rice, Lodato  \& Armitage}{Rice
  et~al.}{2005}]{Rice_et_al_05}
Rice W. K.~M.,  Lodato G.,   Armitage P.~J.,  2005, \mn@doi [MNRAS]
  {10.1111/j.1745-3933.2005.00105.x}, 364, L56

\bibitem[\protect\citeauthoryear{Rix \& Zaritsky}{Rix \&
  Zaritsky}{1995}]{Rix1995}
Rix H.-W.,  Zaritsky D.,  1995, \mn@doi [The Astrophysical Journal]
  {10.1086/175858}, 447, 82

\bibitem[\protect\citeauthoryear{Rosse}{Rosse}{1850}]{Rosse1850}
Rosse E.~O.,  1850, Philosophical Transactions of the Royal Society of London,
  140, 499

\bibitem[\protect\citeauthoryear{Schinnerer et~al.,}{Schinnerer
  et~al.}{2013}]{Schinnerer2013}
Schinnerer E.,  et~al., 2013, \mn@doi [The Astrophysical Journal]
  {10.1088/0004-637X/779/1/42}, 779, 42

\bibitem[\protect\citeauthoryear{Schinnerer et~al.,}{Schinnerer
  et~al.}{2017}]{Schinnerer2017}
Schinnerer E.,  et~al., 2017, \mn@doi [The Astrophysical Journal]
  {10.3847/1538-4357/836/1/62}, 836, 62

\bibitem[\protect\citeauthoryear{Seiden \& Gerola}{Seiden \&
  Gerola}{1979}]{Seiden1979}
Seiden P.~E.,  Gerola H.,  1979, \mn@doi [The Astrophysical Journal]
  {10.1086/157366}, 233, 56

\bibitem[\protect\citeauthoryear{Seigar}{Seigar}{2011}]{Seigar2011}
Seigar M.,  2011, ISRN Astronomy and Astrophysics, 2011

\bibitem[\protect\citeauthoryear{Sellwood \& Carlberg}{Sellwood \&
  Carlberg}{1984}]{Sellwood1984}
Sellwood J.~A.,  Carlberg R.~G.,  1984, \mn@doi [ApJ] {10.1086/162176}, 282, 61

\bibitem[\protect\citeauthoryear{{Sie Kam}, Carignan, Chemin, Amram  \&
  Epinat}{{Sie Kam} et~al.}{2015}]{Kametal2015}
{Sie Kam} Z.,  Carignan C.,  Chemin L.,  Amram P.,   Epinat B.,  2015, MNRAS,
  449, 4048

\bibitem[\protect\citeauthoryear{Springel}{Springel}{2010}]{Springel2010}
Springel V.,  2010, \mn@doi [MNRAS] {10.1111/j.1365-2966.2009.15715.x}, 401,
  791

\bibitem[\protect\citeauthoryear{Teuben}{Teuben}{1995}]{Teuben1995}
Teuben P.,  1995, in Shaw R.,  Panye H.,   Hayes J.,  eds, Astronomical Data
  Analysis Software and Systems IV. Astronomical Society of the Pacific, p.~398

\bibitem[\protect\citeauthoryear{Tobin et~al.,}{Tobin et~al.}{2016}]{Tobin2016}
Tobin J.~J.,  et~al., 2016, \mn@doi [Nature] {10.1038/nature20094}, 538, 483

\bibitem[\protect\citeauthoryear{Vigan et~al.,}{Vigan et~al.}{2017}]{Vigan2017}
Vigan A.,  et~al., 2017, Astronomy {\&} Astrophysics, 603, A3

\bibitem[\protect\citeauthoryear{Zel'dovich}{Zel'dovich}{1970}]{Zeldovich1970}
Zel'dovich Y.~B.,  1970, Astronomy and Astrophysics, 5, 84

\bibitem[\protect\citeauthoryear{Zhu, Dong, Stone  \& Rafikov}{Zhu
  et~al.}{2015}]{Zhu2015}
Zhu Z.,  Dong R.,  Stone J.~M.,   Rafikov R.~R.,  2015, \mn@doi [The
  Astrophysical Journal] {10.1088/0004-637X/813/2/88}, 813, 88

\makeatother
\end{thebibliography}

\label{lastpage}

\end{document}